\documentclass[referee, sn-aps]{sn-jnl}% American Physical Society (APS) Reference Style
\jyear{2022}%

\theoremstyle{thmstyleone}%
\usepackage{caption}
\usepackage{ulem}

\theoremstyle{thmstyletwo}%

\theoremstyle{thmstylethree}%

\newcommand {\nustar} {\textsl{NuSTAR}}

\newcommand {\nicer} {\textsl{NICER}}

\def \src {SGR~1935$+$2154}

\usepackage{color}
\usepackage{threeparttable}
\usepackage{hyperref}
\usepackage{natbib}
\usepackage{multibib}
\usepackage[shortlabels]{enumitem}

\usepackage{lscape}

\def\revi#1{\textcolor{black}{#1}}

\begin{document}

\title[SGR 1935 FRB and Glitch]{Rapid spin changes around a magnetar fast radio burst}

\maketitle
\begin{center}
Chin-Ping~Hu$^{1,2}$, Takuto~Narita$^3$, Teruaki~Enoto$^{3,2}$, George~Younes$^{4}$, Zorawar~Wadiasingh$^{4,5,6}$, Matthew~G. Baring$^{7}$, Wynn~C.~G.~Ho$^{8}$, Sebastien~Guillot$^9$, Paul~S.~Ray$^{10}$, Tolga~G\"uver$^{11,12}$, Kaustubh~Rajwade$^{13}$, Zaven~Arzoumanian$^{4}$, Chryssa~Kouveliotou$^{14}$, Alice~K.~Harding$^{15}$, and Keith~C.~Gendreau$^4$\\[0.5cm]
\end{center}

\begin{enumerate}[label={$^{\arabic*}$}]
\item Department of Physics, National Changhua University of Education, No.1, Jinde Rd., Changhua City, 50007, Taiwan
\item Extreme Natural Phenomena RIKEN Hakubi Research Team, Cluster of Pioneering Research, RIKEN, 2-1 Hirosawa, Wako, 351-0198, Saitama, Japan
\item Department of Physics, Graduate School of Science,  Kyoto University, Kitashirakawa, Sakyo-ku, Kyoto, 606-8502, Kyoto, Japan
\item Astrophysics Science Division, NASA Goddard Space Flight Center, 8800 Greenbelt Road, Greenbelt, MD, 20771, USA
\item Department of Astronomy, University of Maryland College Park, 4296 Stadium Dr., PSC, College Park, MD, 20742, USA
\item Center for Research and Exploration in Space Science and Technology, NASA/GSFC, 8800 Greenbelt Road, Greenbelt, MD, 20771, USA
\item Department of Physics and Astronomy, Rice University, 6100 Main Street, Houston, Texas, 77251-1892, USA
\item Department of Physics and Astronomy, Haverford College, 370 Lancaster Avenue, Haverford, PA, 19041, USA
\item Institut de Recherche en Astrophysique et Plan\'{e}tologie, UPS-OMP, CNRS, CNES, 9 avenue du Colonel Roche, BP 44346, Toulouse Cedex 4, 31028, France
\item Space Science Division, U.S. Naval Research Laboratory, 4555 Overlook Ave., SW, Washington, 20375, DC, USA
\item Science Faculty, Department of Astronomy and Space Sciences, Istanbul University, Beyaz\i t, 34119, Istanbul, T\"urkiye
\item Observatory Research and Application Center, Istanbul University, Beyaz\i t, 34119, Istanbul, T\"urkiye
\item ASTRON, the Netherlands Institute for Radio Astronomy, Oude Hoogeveensedijk 4, 7991 PD, Dwingeloo, The Netherlands
\item The George Washington University, 725 21$^{\textrm{st}}$ street, NW, 20052, Washington, DC, USA
\item Theoretical Division, Los Alamos National Laboratory, Los Alamos, NM 87545, USA
\end{enumerate}

%\textcolor{cyan}{(cphu) I rewrote the opening paragraph. \mgbc{tweaked slightly by MGB} It's now roughly 200 words.}
\vspace{1cm}

\revi{Magnetars are neutron stars with extremely high magnetic fields ($\gtrsim10^{14}$~Gauss) that exhibit various X-ray phenomena such as sporadic sub-second bursts, long-term persistent flux enhancements, and variable rates of rotation period change \citep{Kouveliotou1998, KaspiBeloborodov2017}.  In 2020, a fast radio burst (FRB), akin to cosmological millisecond-duration radio bursts, was detected from the Galactic magnetar \src~\cite{CHIME2020_SGR1935, MereghettiSF2020, BochenekRB2020}, confirming the long-suspected association between some FRBs and magnetars.  However, the mechanism for FRB generation in magnetars remains unclear. Here we report the X-ray discovery of an unprecedented double glitch in \src\ within a time interval of approximately nine hours, bracketing an FRB that occurred on October 14, 2022 \citep{Dong_CHIME_2022,MaanLS2022}. Each glitch involved a significant increase in the magnetar's spin frequency, being among the largest abrupt changes in neutron star rotation \cite{EspinozaLS2011,2013MNRAS.429..688Y,2022MNRAS.510.4049B} ever observed. Between the glitches, the magnetar exhibited a rapid spin-down phase, accompanied by a profound increase and subsequent decline in its persistent X-ray emission and burst rate. We postulate that a strong, ephemeral, magnetospheric wind \citep{YounesBH2023} provides the torque that rapidly slows the star's rotation. The trigger for the first glitch couples the star's crust to its magnetosphere, enhances the various X-ray signals, and spawns the wind that alters magnetospheric conditions that might produce the FRB. }

\src, the most active Galactic magnetar of the last decade with a spin period of 3.25 s and a dipole (polar) magnetic field strength of $4\times 10^{14}$~G \cite{Israel+16}, has shown several major X-ray and radio outbursts \cite{Younes+17,YounesGK2020}.
The most recent outburst episode of \src\ occurred in October 2022 and lasted several days during which it emitted hundreds of short X-ray bursts \cite{MereghettiGF2022, Palm2022, YounesEH2022} (see below).
At Coordinated Universal Time (UTC) 19:21:47 (topocentric time of the Canadian Hydrogen Intensity Mapping Experiment, hereafter CHIME) on October 14, 2022, this magnetar emitted an FRB with multiple radio peaks measured with CHIME and the Robert C.~Byrd Green Bank Telescope (GBT) \cite{Dong_CHIME_2022, MaanLS2022}.

\begin{figure}[h]%
\centering
\includegraphics[width=0.9\textwidth]{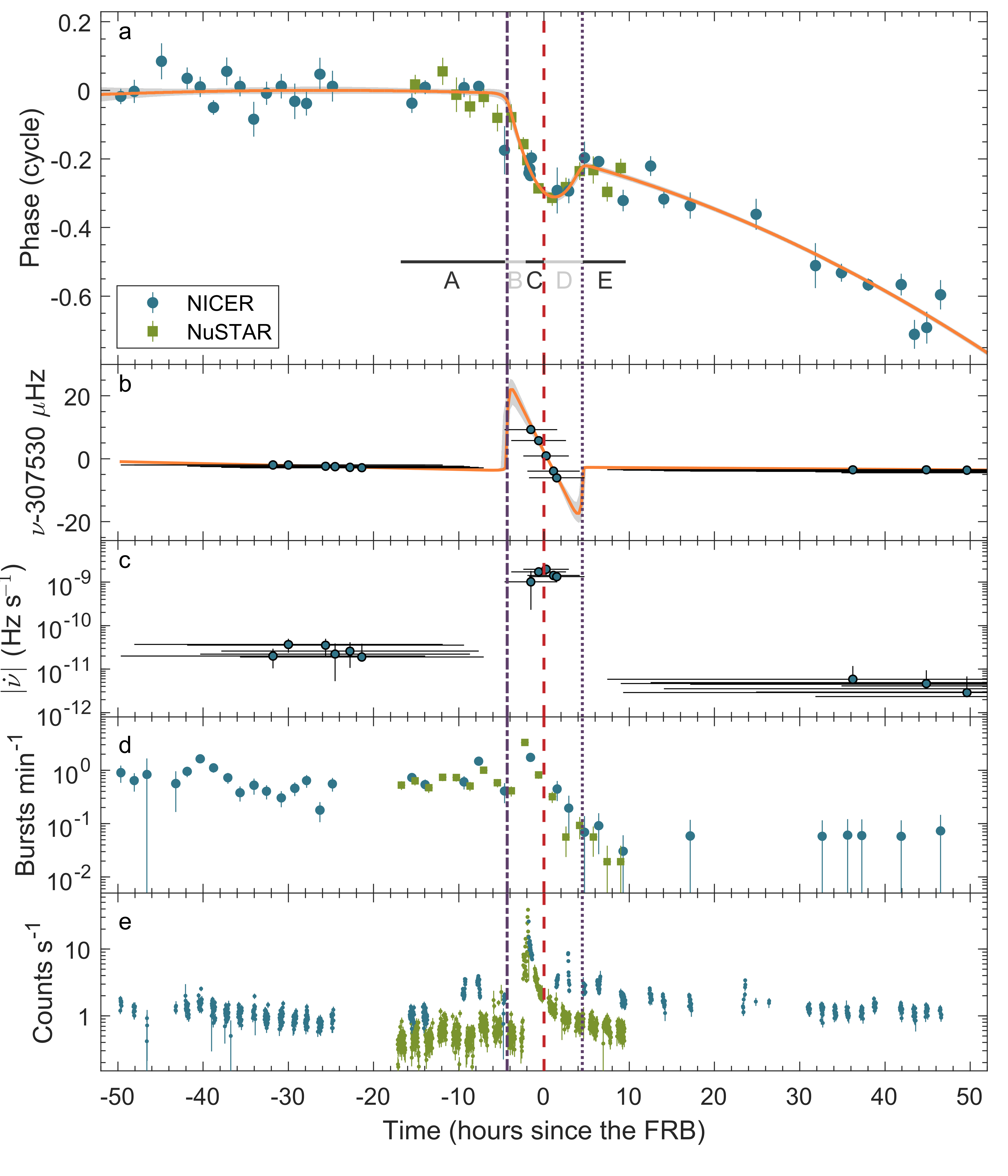}
\caption{\revi{\bf Evolution of pulse timing, short X-ray bursts, and flux near the double-glitch epoch.} The time zero epoch, marked by a vertical red dashed line, is defined as the occurrence time of the CHIME FRB. The dashed-dotted and dotted vertical lines denote the times of the first glitch and second glitch, respectively.  Panel a shows the evolution of the pulse arrival phase. The orange line denotes the best-fit timing solution and the gray-filled area is the 1-$\sigma$ confidence interval obtained with MCMC simulations. Black and gray horizontal lines (labeled A through E) denote the time intervals of spectral analysis.  Panel b describes the frequency evolution derived from the best-fit timing model (orange line) and the corresponding 1-$\sigma$ confidence interval. The data points are best-fit frequencies obtained from 15 consecutive arrival phases with a moving window of one data point. Panel c is the $\dot{\nu}$ evolution obtained using the same procedure as in panel b. Panel d shows the evolution of the burst occurrence rate, where blue circles denote the values obtained with \nicer\ and green squares denote those obtained with \nustar.  Panel e shows the count rate of the persistent emission after removing the bursts. }\label{fig:fig1_toa_analysis}
\end{figure}

Alerted prior to this activity \cite{MereghettiGF2022}, we initiated a series of Neutron Star Interior Composition Explorer (\nicer) with 75~ks exposure and Nuclear Spectroscopic Telescope Array (\nustar) X-ray observations with 96~ks exposure starting from 17:32:40 UTC on October 12 until November 06, 2022 (Extended Data Table \ref{tab:observation_log}).
%, 2022,  with a total exposure time of 75~ks for \nicer\ and 96~ks for \nustar\ as of  
Hereafter, we use the elapsed time $t$ since the FRB detection at barycentric corrected  Modified Julian Date (MJD)~59866.80817034.
Our high-cadence X-ray observations attained 67\% temporal coverage bracketing the FRB, $-17<t<10$ hours, but did not cover the FRB itself due to Earth occultation. 
Nevertheless, this cadence enabled an unprecedented investigation of the spin-frequency evolution of the magnetar around the FRB.
%\sout{time of the radio activity}.
%\sout{high-cadence X-ray observations} 
After removing short bursts with \revi{timescales from milliseconds to around ten seconds}, the stellar rotational frequency $\nu$ can be revealed by pulsed thermal emission from the stellar surface using \nicer\ in 2--8~keV and \nustar\ in 3--8~keV. 
Initially, the spin behavior exhibited a spin-down rate of $\dot{\nu}=-1.7(6)\times10^{-11}$ Hz s$^{-1}$ (Figure \ref{fig:fig1_toa_analysis}a), with parentheses indicating the 1$\sigma$ uncertainty on the last digit.
%\sout{frequency $\nu=0.3075277(1)$ Hz and its time derivative,} 
%\sout{at a time-zero reference epoch of $T_0=-27.12$ hours} 
%by pulsed thermal emission from the stellar surface

A precise timing model was then derived by phase-coherent analysis, i.e., tracking the X-ray pulse times-of-arrival (TOAs) with a predictive model and optimizing its parameters \cite{LivingstoneRC2009}.  
%\sout{The TOAs were computed using the unbinned maximum-likelihood method.}
In the phase evolution (Figure \ref{fig:fig1_toa_analysis}a and Extended Data Figure \ref{fig:toa_analysis_supl1}), a concave trend is observed in the range of $-4.4\lesssim t \lesssim 4.4$ hours, followed by a subsequent long-term convex quadratic trend.
%These data points are satisfactorily fitted with a model that includes two spin-up glitches, with the first one occurring at $t_{g1}=-4.4_{-0.5}^{+0.3}$ hours (MJD $59866.63_{-0.02}^{+0.01}$) and the second one at $t_{g2}=4.4_{-0.5}^{+0.4}$ hours (MJD $59866.99\pm0.02$). 
These data points are satisfactorily fitted by a model (see Methods) incorporating two spin-up glitches at $t_{g1}=-4.4_{-0.5}^{+0.3}$ hours (MJD $59866.63_{-0.02}^{+0.01}$) and $t_{g2}=4.4_{-0.5}^{+0.4}$ hours (MJD $59866.99\pm0.02$). 
With an uncertainty of approximately $\pm30$ minutes, these constitute the best-constrained glitch epochs ever observed in magnetars and pulsars at high energies \cite{EspinozaLS2011, DibK2014}. 
We attempted to fit the TOAs using a high-order polynomial but the resulting reduced $\chi^2$ value was much higher than that of the best-fit two-glitch model (see Methods). 
%\revi{and transient timing noise}

\begin{figure}[h]%
\centering
\includegraphics[width=0.9\textwidth]{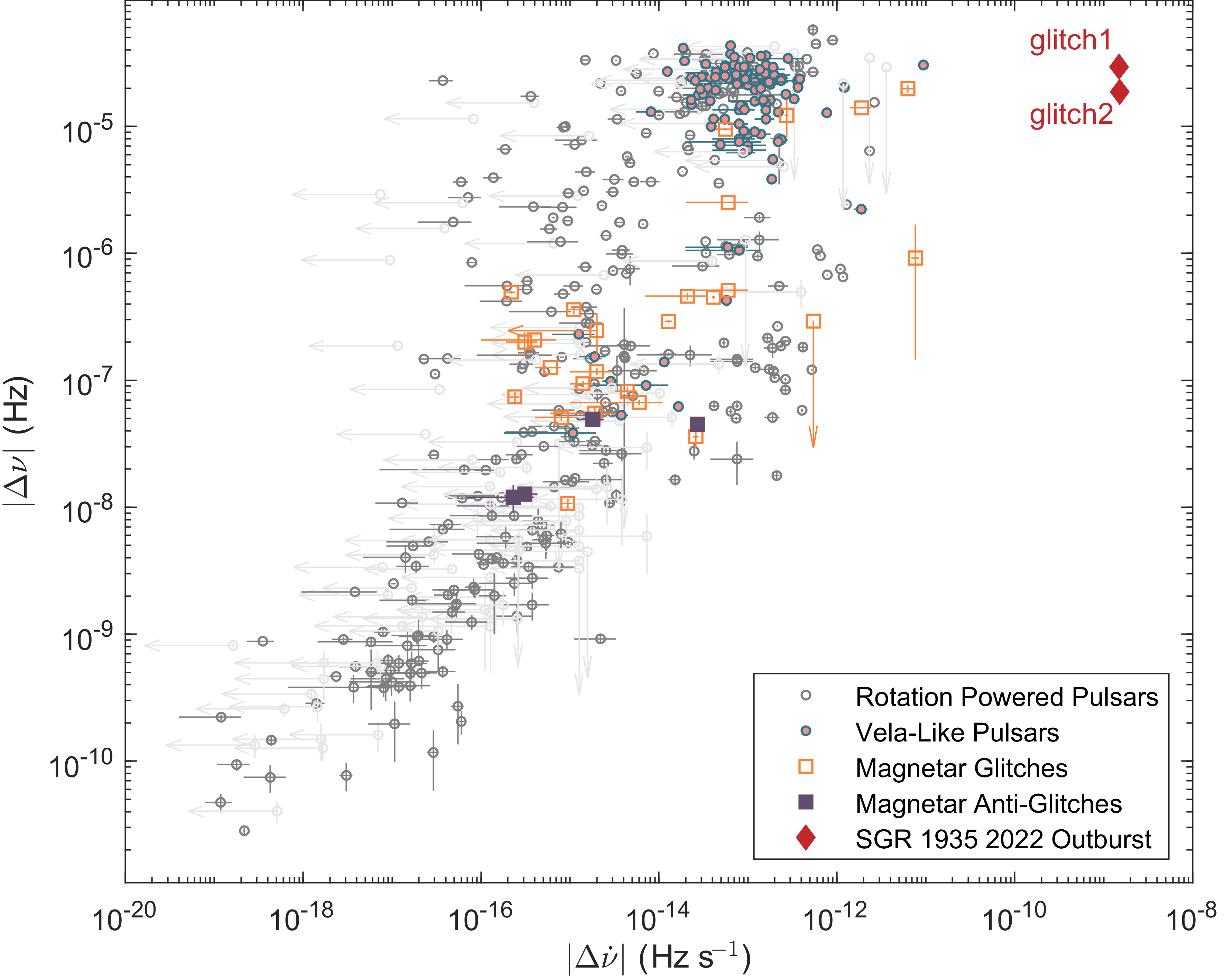}
\caption{\revi{\bf Pulse frequency discontinuities for known pulsar glitches.} Glitch size distribution of previously reported rotation-powered pulsars (circles) and magnetars (squares)~\cite{EspinozaLS2011, 2022MNRAS.510.4049B} in the phase space of changes of stellar spin frequencies (ordinate) and their derivatives (abscissa) between glitches. The data are shown as absolute values. \revi{The glitches of rotation-powered pulsars are bimodally distributed, where glitches of Vela-like pulsars (blue circles filled with pink) mainly lie on the top of the figure, and glitches of other pulsars form a linear trend with scatter \cite{EspinozaLS2011}. } The two magnetar glitches observed in this work are indicated by red diamonds. \revi{They have extremely large changes in $\vert \Delta {\dot \nu}\vert$ compared to all pulsars, and $\vert \Delta \nu\vert$ comparable to those of Vela-like pulsars. These two glitches lie on the extension of the linear trend of pulsar glitches.}
\label{fig:glitch_size_distribution}
%Glitch size distribution of all pulsars and magnetars. We included candidate anti-glitches of magnetars and plotted their absolute values in the glitch size $\Delta\nu$ as filled squares.  }\label{fig:glitch_size_distribution
}
\end{figure}

At the first glitch, the frequency jumps by $\Delta\nu_1=3.0(3)\times10^{-5}$ Hz, corresponding to a fractional change of $\Delta\nu_1/\nu=1.0(1)\times10^{-4}$.
The spin-down rate $\dot{\nu}$ also changed by $\Delta\dot{\nu}_1=-1.5(3)\times10^{-9}$ Hz s$^{-1}$. 
The second glitch had a smaller $\Delta\nu_2=1.9(3)\times10^{-5}$~Hz ($\Delta\nu_2/\nu=6(1)\times10^{-5}$) with $\Delta\dot{\nu}_2=1.5(3)\times10^{-9}$~Hz~s$^{-1}$, \revi{thereby compensating for the excess recovery that occurred between the two glitches.}
Uncertainties are obtained through Markov-Chain Monte Carlo (MCMC) simulations (see Methods).  
Figure \ref{fig:glitch_size_distribution} compares the detected glitches with previously measured magnitudes in magnetars and pulsars \cite{EspinozaLS2011, HuN2019, 2022MNRAS.510.4049B, YounesBH2023, GeYL2023}, demonstrating that these two glitches are among the largest ever observed.  
\revi{The rotational energy increments associated with these two glitches are $3.9\times10^{41}$~erg and $2.6\times10^{41}$~erg, respectively (see Methods).}

Panels b and c of Figure~\ref{fig:fig1_toa_analysis} show the frequency evolution and its derivative using the two-glitch model.
Additionally, we obtained the local spin ephemeris by allowing overlaps among consecutive TOAs and avoiding time intervals that cross the glitch epochs. 
%---these data points represent a boxcar moving-average smoothing of the true evolution with adaptive window sizes utilized to accommodate the non-uniform sampling of the TOAs. 
The $\vert {\dot \nu}\vert$ between the two glitches increased by a factor of $\approx100$ compared to the pre-glitch value, signaling a rapid spin-down bracketing the time of the FRB.

To investigate the evolution of short burst occurrence rate (Figure \ref{fig:fig1_toa_analysis}d) in this episode, we applied the Bayesian-block technique to search for burst candidates using the photon events collected by \nicer\ and \nustar\ \revi{(see Methods)}.
% in the 2--8 keV and 3--79 keV energy bands, respectively.
%The evolution of the burst occurrence rate is shown in Figure \ref{fig:fig1_toa_analysis}d.
%We identified 378 and 720 short burst candidates from the \nicer\ and \nustar\ data (see Methods)
%Figure \ref{fig:fig1_toa_analysis}d shows the burst occurrence rate, which was calculated by dividing the number of detected bursts by the successive good time interval (GTI) duration in each satellite orbit.
The burst-active epoch lasted for at least two days, during which the rate remained at approximately one burst per minute. 
Subsequently, at $t\approx -2$ hours, the rate peaked at more than four bursts per minute and subsequently dropped to below 0.05 bursts per minute in the next ten hours, \revi{being less than 10\% of the pre-glitch level at $t_{g2}$.}
This declining phase is contemporaneous with the occurrence of the FRB.  
Such interplay between the X-ray bursts and FRB is similar to the behavior seen during the previous outburst of \src\ observed in 2020 \cite{YounesGK2020, YounesBK2021}.

After excising these burst intervals, we extracted the X-ray photons and plotted the 64-s binned count rates of the persistent X-ray emission (Figure \ref{fig:fig1_toa_analysis}e).  
The \nicer\ and \nustar\ count rates at the early outburst phase ($-51<t<-2.5$ hours) were approximately 1.0 and 0.5 count~s$^{-1}$.
% in the 2--8~keV and 3--79~keV bands, respectively
The persistent X-ray flux remained constant until a sudden increase at $t=-2.5$ hours, reaching a \nustar\ count rate of over 50 count~s$^{-1}$ within one minute. 
Subsequently, the count rate declined to approximately 0.5 count~s$^{-1}$ within ten hours.  \revi{We also note that the FRB is phase-correlated with the 20-79 keV pulse peak, which exhibits a 0.5 phase offset relative to the soft X-ray pulse profile (see Methods).}

% The energy bands for our timing analysis were 2--8~keV and 3--8~keV recorded with \nicer\ and \nustar, respectively. 
% These energy ranges are dominated by pulsed thermal emission from the stellar surface and provide an excellent proxy for measuring rotation.
% , and a mini outburst with hundred of peaks lasting a few minutes
% Twenty-five minutes after the rising edge, the \nustar\ count rate further increased to $>100$ count~s$^{-1}$ and exhibited hundreds of burst-like peaks. 
%On October 10, 2022, \src\ entered a new burst active state starting with several X-ray short bursts. 
% These data were processed using the standard data analysis procedures of each satellite (see Methods). 
% The evolutionary trend of the burst rate and the persistent emission are consistent with each other when taking into account the bin size of the burst rate. 

\begin{figure}[h]%
\centering
%\includegraphics[width=0.9\textwidth]{figure/tmp_sgr1935_spectra_v230203.jpg}

%comment
%\includegraphics[width=1.0\textwidth]{figure/fig_sgr1935_eeuf_v230213.pdf}
\includegraphics[width=1.0\textwidth]{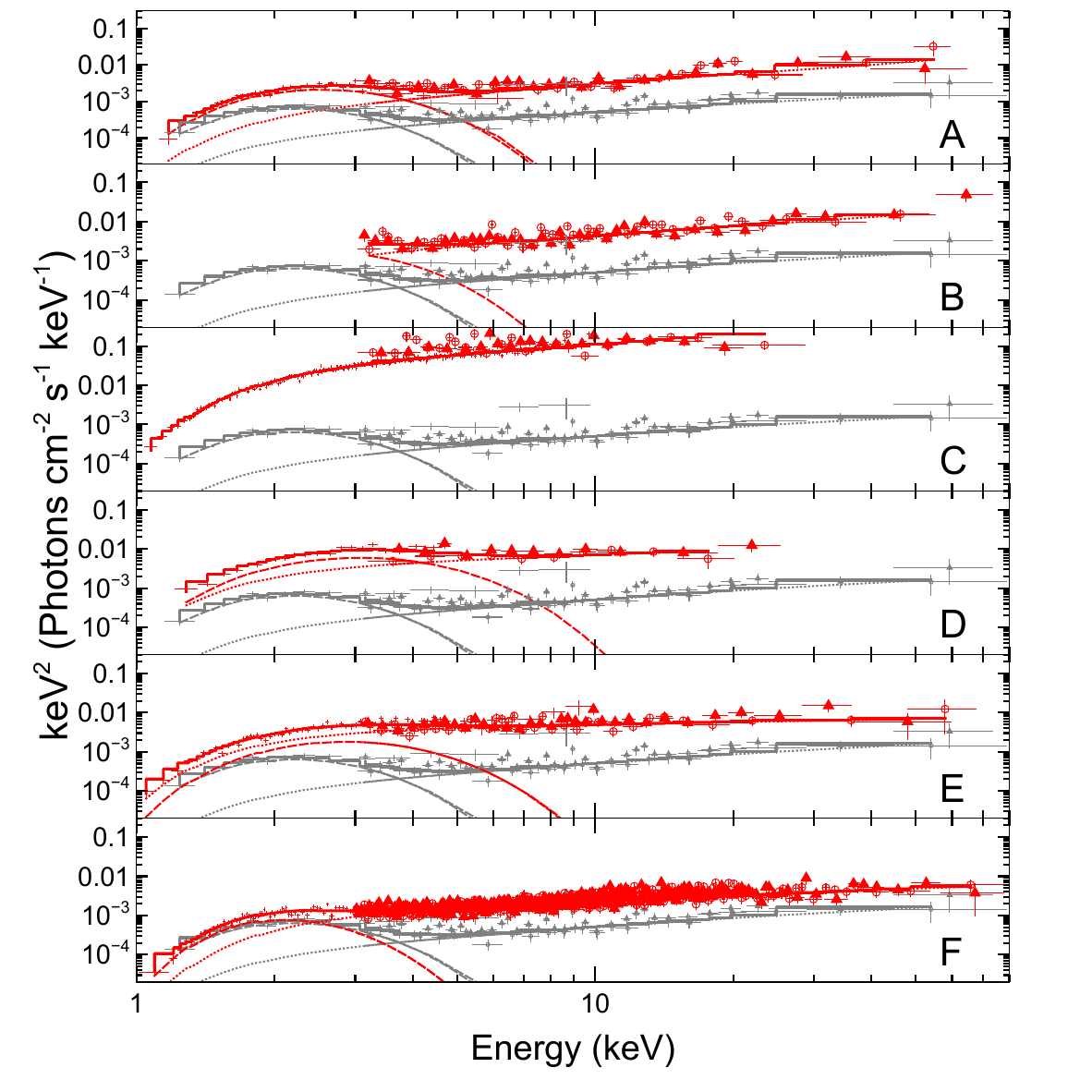}
\caption{\revi{\bf X-ray spectral variations near the double-glitch epoch.} X-ray spectral evolution of the persistent emission (red) of SGR~1935+2154 after removing short bursts, obtained with \nicer\ and \nustar\ in time epochs A to F (see panel a of Figure~\ref{fig:fig1_toa_analysis}). 
\nicer\ spectra are shown in crosses, and \nustar\ FPMA and FPMB spectra are shown in circles and triangles, respectively.
As a reference, we compare each of these active-state spectra with one quiescent spectrum (gray) obtained on October 16, 2020, two years before the present FRB.
%The left and right panels correspond with the radiation peak's rising and decaying phases.
The spectra of epochs A, B, D, E, F, and the quiescent state are fit using a blackbody plus a power law with absorption in the interstellar medium. The spectra in epoch C are approximated with a power law. The spectral components, including photo-absorption, are plotted using dashed (blackbody) and dotted (power law) lines. 
%Photo-absorption is included in the spectra. 
}\label{fig:fig_spectrum}
\end{figure}

We also investigated the \nicer\ and \nustar\ X-ray spectral behavior contemporaneous to the FRB (Figure~\ref{fig:fig_spectrum}). 
We divided the data into six time periods; (A) the early stage $t<t_{g1}$, (B) from $t_{g1}<t<-2.5$ hours, the peak of the X-ray emission (C) $-2.5<t<-0.2$~hours before the FRB, (D) $0.5<t<t_{g2}$, (E) decaying phase at $t_{g2}<t<10$~hours (Figure \ref{fig:fig1_toa_analysis}a), and (F) the returning to quiescence at $11.5$~\revi{days} $<t<12.2$~days (Extended Data Table \ref{tab:observation_infomation}). The best-fit spectral parameters are summarized in Extended Data Table \ref{tab:best_fit_spectral_parameter}.
The quiescent spectrum is well characterized by two components, as reported in other magnetars \citep{EnotoSK2017}; a blackbody from a hotspot on the stellar surface with a temperature of $kT=0.42^{+0.04}_{-0.03}$~keV and a radius $R=1.8^{+0.5}_{-0.3}$~km when assuming a fiducial distance $d=6.6$~kpc \cite{Zhou+20}, and a non-thermal power-law component from the magnetosphere with a photon index of $\Gamma=1.2\pm 0.1$. 

%We carried out a series of \nicer\ observations with a total exposure time of 7.4 ks, and two \nustar\ observations with a total exposure time of 9.6 ks before November 06, 2022 (see Extended Data Table \ref{tab:observation_log} for detailed information). 
%, of which burst count rates exceeded XX and XX, respectively (see Method).

%We then remove these bursts and obtain the flux evolution of the persistent emission by binning the non-burst photons into a light curve with a bin size of 64 seconds (see Figure \ref{fig:fig1_toa_analysis} (b)). A flare with a rapidly rising edge can be seen about three hours before the FRB, and then the flux decays exponentially. The burst occurrence rate shows a similar evolutionary trend. The peak of the burst occurrence is consistent with the peak of the persistent flux after taking the bin size into account though the burst occurrence rate may be underestimated at the peak (see Methods for detail).  

%The high cadence \nicer\ and \nustar\ monitoring around the FRB-related glitches allowed us to investigate the spectral evolution of the persistent X-ray emission as summarized in Figure~\ref{fig:fig_spectrum}.

At epoch A, the persistent X-ray flux increased by a factor of five compared to the quiescent flux level. The spectrum is modeled by the same two-component model but with a hotter blackbody temperature at $kT=0.51\pm 0.03$~keV with $R=2.1\pm{0.2}$~km and $\Gamma=1.1\pm0.1$. At epoch B, no significant flux change was observed. Subsequently, the flux increased by a factor of $>10$ at 2.5 hours before the FRB (epoch C), with the power-law component dominating the X-ray spectrum (see detailed analysis in Methods).  After the FRB (epoch D), the flux gradually decreased and, just after the second glitch (epoch E), returned to a similar value as that of epoch A with a soft $\Gamma$ of 1.6--1.8.
At epoch F, the flux dropped to half of the level in epoch E, yet was still 3 times higher than the quiescent flux.

%The spectral analysis of burst emission is described in the Method section (Extended Data Figure~\ref{fig:spectral_fit_results_burst}). 
%accompanied by the elevated bursting activity
% , including both the blackbody and the power-law components
%The spectrum is again approximated by two components but with a soft photon index of $\Gamma=1.6-1.8$.

%The spectral analysis allows an estimation of the energetics of radiative output through bursts and persistent emission around the FRB. 
The total isotropic-equivalent radiative energy of the persistent emission is estimated as $5.0 \times10^{40}~(d/6.6~\rm{kpc})^2$~erg between the two glitches. The burst emission at epochs B, C, and D has a total energy of {$2.3~\times 10^{40}~(d/6.6~\rm{kpc})^2$~erg} when fitted with a power-law spectrum. This X-ray radiative output, $7.3~\times 10^{40}~(d/6.6~\rm{kpc})^2$~erg, is one order of magnitude lower than the spin-down inferred energy but two orders of magnitude higher than the normal quiescent energy of $\lesssim 10^{39}~(d/6.6~\rm{kpc})^2$ erg within the same time span (Extended Data Table \ref{tab:best_fit_spectral_parameter}).  

We outline one possible picture of the unique behavior detailed above, which may provide clues to the environment that permits or triggers the observed FRB, and radio bursts from \src\ in general.  
Owing to the absence of major changes in the count rate, pulse profile, and spectra around $t_{g1}$, the first glitch was probably triggered by a mechanism internal to the magnetar.  Spin-up glitches in normal pulsars \cite{EspinozaLS2011,2013MNRAS.429..688Y,2022MNRAS.510.4049B} and some magnetars \cite{DibK2014} are thought to be due to angular momentum transfer from superfluid neutrons in the rapidly rotating inner crust (and possibly core) to the other parts of the star \cite{Anderson1975}, with the latter being slowed over long times by magnetospheric torques.  The rapid spin-down after the first glitch quickly regenerated the spin lag between the superfluid and the rest of the star, leading to the second glitch.  
\revi{Using angular momentum conservation arguments (see Methods)}, the short time interval of 8.8~hours between glitches suggests a \revi{superfluid component of several tens of percent}.  This implies that a large fraction of the core, along with most free neutrons in the inner crust, is in a superfluid state.

Long-term magnetic activity in the outer layers can provide heat or excite oscillations that can induce the unpinning of superfluid vortices in deeper layers \cite{Link1996, EichlerS2010}.   The motion of a large number of superfluid vortices could move superconducting flux tubes in the core.  The latter would then change the surface layer magnetic field geometry, stress the crust, and heat its outer layers, thereby possibly producing the increased persistent emission and burst rate a few hours after the first glitch (see Methods).  These stresses rupture the crust near the magnetic poles where the field is strongest. The associated heating to super-Eddington temperatures can lead to the expulsion of large amounts of \revi{ion-rich plasma (``volcanism'' into the magnetosphere), forming an optically thick and collimated relativistic wind that torques the star}.  
%\sout{This ``volcanism'' generates an optically thick, highly collimated, mildly relativistic wind that flows outwards to escape the magnetosphere along open field lines \citep{YounesBH2023}.  This wind is ephemeral and torques the rotating magnetar by extracting angular momentum.}  
This ephemeral wind can naturally explain the rapid spin-down epoch between the glitches, provided the cumulative mass expelled is sufficient (see Methods). 
%\sout{The timing model suggests that the wind abates somewhat contemporaneously with the second glitch, by which time it has re-established the rotational lag between the superfluid component and the rest of the star, leading to the second glitch.  Such seismic motions and the quasi-polar wind will reconfigure the field at the surface and in the magnetosphere.} 
The wind twists field lines and increases the magnetic field energy density \citep{TLK-2002-ApJ,Hu-2022-ApJ} %,Mahlmann2022ApJ} 
in the magnetosphere over the poles.  As it abates, it untwists the field and clears the magnetosphere of current, perhaps seeding the decline of the X-ray flux and burst rate before the radio burst (see Figure \ref{fig:fig1_toa_analysis}d, \ref{fig:fig1_toa_analysis}e). 
%\sout{The operation of magnetic pair cascades, thought to underlie radio emission \citep{2013MNRAS.429...20T}, and its interplay with photon splitting, is generally sensitive to the location and strength of field twists \citep{Hu-2022-ApJ}. These cascades also supply the plasma demanded by field twists or bursts driven by the crust.} 

A possible scenario for the FRB is the ignition of pair cascades by a magnetar short burst as the magnetosphere evolves to a more charge-starved state \cite{WadiasinghT2019}. Such phenomenology is generally consistent with the events around the April 2020 FRB, which occurred many hours into the waning phase of intense burst activity \cite{YounesGK2020} and the post-anti-glitch radio burst in October 2020 \citep{YounesBH2023}. 
\revi{Future high-cadence observations of \src\ and other magnetars in X-rays, like those presented here and in conjunction with radio monitoring, will help identify the conditions required for generating FRBs.}

%%%%%%%%%%%%%%% Figure 1: TOA Analysis %%%%%%%%%%%%%%%%%%%%%%%%

%%%%%%%%%%%%%%% Glitch Size %%%%%%%%%%%%%%%%%%%%%%%%
%%%%%%%%%%%%%%% Glitch Size%%%%%%%%%%%%%%%%%%%%%%%%

%%%%%%%%%%%%%%% Spectral Analysis %%%%%%%%%%%%%%%%%%%%%%%%
%Errors are statistical 1$\sigma$. $
%%%%%%%%%%%%%%% Spectral Analysis %%%%%%%%%%%%%%%%%%%%%%%%

%In case of double column layout, the above format puts figure captions/images to single column width. To get spanned images, we need to provide \verb+\begin{figure*}+ \verb+...+ \verb+\end{figure*}+.

%For sample purpose, we have included the width of images in the optional argument of \verb+\includegraphics+ tag. Please ignore this. 

\backmatter

\clearpage

\renewcommand\tablename{Extended Data Table}
\setcounter{figure}{0} 
\renewcommand\figurename{Extended Data Figure}
\section*{Methods}
%\begin{methods}

\noindent \textbf{\nicer\ data reduction.} 
%We began monitoring \src\ using \nicer\ on October 12, 2022, at 17:32:40 UTC. 
%We collected 19 observations with a total exposure time of 75~ks (see Extended Data Table \ref{tab:observation_log}).
\nicer\ is a non-imaging X-ray observatory with a large collecting area of 1900 cm$^2$ at 1.5 keV and an excellent time resolution of $<300$~nanoseconds aboard the International Space Station \cite{GendreauAA2016}. It consists of 56 co-aligned X-ray concentrating optics and associated silicon-drift detectors, of which 52 are in operation. 
%The large collecting area of 1900 cm$^2$ at 1.5 keV and the excellent time resolution of $<300$~nanoseconds.
%make \nicer\ an ideal instrument for studying short bursts and pulsations of magnetars.    
We reprocessed the \nicer\ data using NICERDAS version 9, part of the HEASOFT version 6.30.1 package and the calibration database version 20210720. 
The unfiltered data were calibrated and screened using the nicerl2 pipeline with default settings, resulting in clean scientific event lists. 
\revi{We further removed a few background flares near GTI boundaries.}
We corrected the arrival time of each photon to the barycenter of the solar system using the barycorr command with the JPL solar-system ephemeris DE405. 
Photons with energies below 2 keV, due to significant absorption, and above 8 keV, due to reduced \nicer\ sensitivity, were excluded from the analysis.
A summary of the \nicer\ data used in this research can be found in Extended Data Table \ref{tab:observation_log}. 

\noindent \textbf{\nustar\ data reduction.} 
%\src\ was observed with NuSTAR\cite{HarrisonCC2013} on October 14 at 02:01:09 UTC (ObsID 80802317002) and October 16 at 08:35:11 UTC (ObsID 80802317004). 
\nustar\ consists of two identical focal plane modules (FPM), FPMA and FPMB. \revi{Each module is equipped with} conical Wolter-I optics that achieve a point spread function with a half-power diameter of around 60 arcseconds. 
%, or a full width at half maximum of $\sim$18 arcseconds
Each FPM houses a detector with an energy range of 3--79 keV, a field of view of $12'\times12'$, an energy resolution of 0.4 keV at 10 keV, and a timing resolution of 65 microseconds\cite{BachettiMG2021}. The data were reduced using NUSTARDAS version 2.1.2 with the calibration database version 20211020. 
We performed data calibration and screening using the nupipeline script. 
Images, event lists, light curves, and spectra were extracted using nuproducts. 
The source photons were extracted from a circular region with a radius of 0.85~arcmin centered on the position of \src. 
This guarantees that around 85\% of \revi{the} source X-ray photons were encircled. 
%Background regions were selected from a source-free part of the imaged field in the same observation. 
We extracted the background spectra \revi{from a circular source-free region within} a radius of 2.5 arcmin centered on \revi{R.A.~} 19:34:46.6029 \revi{and Decl.} +21:46:55.710 \revi{in the J2000 reference frame}.
%\src\ was burst-active during the observation. 
%The FRB related X-ray burst, however, occurs 12 minutes of the end of a good time interval (GTI). 
%\textcolor{cyan}{(exposure)}.

\noindent \textbf{Burst identification:}
We used the Bayesian block technique\footnote{\href{https://docs.astropy.org/en/stable/api/astropy.stats.bayesian_blocks.html}{https://docs.astropy.org/en/stable/api/astropy.stats.bayesian\_blocks.html}}, which is designed to detect local variability by dividing data into variable-sized time blocks, to search for significant flux changes~\citep{ScargleNJ2013}.
The search energy ranges for the \nicer\ and \nustar\ events were 2--8~keV and 3--79~keV, respectively. 
The prior was computed with a false alarm probability of $p_0=0.01$. 
We first identified blocks with a duration shorter than 4 seconds, and then merged nearby blocks as bright and long bursts are typically identified as several consecutive blocks.  
%Among the simultaneous \nicer\ and \nustar\ good time intervals, 20 bursts were captured simultaneously. We have plotted the light curves of the four brightest bursts in Extended Data Figure \ref{fig:simultaneous_bursts}. The light curves are binned into a resolution of 20 milliseconds. These bright bursts have a relatively long duration of $\gtrsim1$ seconds and often show multiple peaks. 
Next, we assessed the likelihood of the total counts in each block being a Poisson random fluctuation around the count rate present approximately 1~second before and after the block.
We considered all bursts with a probability lower than $3\times 10^{-7}$ (corresponding to a detection significance of 5$\sigma$) to be confirmed bursts and marked the rest as burst candidates.
We removed all the confirmed bursts and burst candidates from the analysis of the persistent count rate and rotation frequency.
In addition, \nustar\ captured a flare with multiple peaks that occurred approximately two hours before the time of the FRB emission. 
The persistent emission during the peak of the flare increased by two orders of magnitude and shows multiple peaks, making it challenging to identify individual bursts. 
We do not attempt to identify the numerous small peaks within the $200$-second long flare. 
\revi{We conservatively count it as a single burst and remove it from further timing analysis since the persistent emission between peaks may be heavily contaminated.}
Therefore, the burst rate at the flux peak in Figure \ref{fig:fig1_toa_analysis}d may be underestimated.
A total of 368 \revi{short} burst candidates were identified in the \nicer\ data set and 720 in the \nustar\ data set. 
%\revi{The detection significance suggests that $\lesssim 5$ detected bursts could be spurious.} 
%We exclude the entire flare peak from our timing analysis. 
Before the FRB, \src\ exhibited a high burst rate of $\gtrsim 1$ \revi{short} bursts per minute, which then dropped to $\lesssim 0.1$ bursts per minute within a time scale of $\sim$5 hours.

\noindent \textbf{Timing analysis and glitch model:}
%The high observational cadence between October 12 and November 6 enabled us to accurately trace the arrival time of the pulses after removing burst candidates. 
We first collected X-ray photons detected with \nicer\ before the FRB and performed a two-dimensional $Z^2$-test with two harmonics (${\rm n}=2$) \cite{BuccheriBB1983}. 
%The native frequency resolution of the data sets with a time span of $T$ can be estimated as $\Delta\nu=1/T$ and $\Delta\dot{\nu}=2/T^2$.
%We began by setting an oversampling rate of 20, i.e., 20 samples within a peak of both $\nu$ and $\dot{\nu}$ dimensions. 
%We searched for the candidate solution among trial $\nu$-$\dot{\nu}$ pairs that result in the highest $Z_2^2$ value. 
%Afterward, we increased the oversampling rate to 1000 and searched for the precise solution near the candidate. 
As the time span increased, we stopped the $Z_2^2$ search when the spin-down rate could be significantly determined, which was $\nu=0.3075277(1)$ Hz and $\dot{\nu}-2.0(6)\times10^{-11}$ Hz s$^{-1}$ at a time zero epoch of MJD 59865.678 (TDB). 
This corresponds to a spin period derivative of $\dot{P}=2.1(6)\times10^{-10}$~s\,s$^{-1}$, an order of magnitude higher than the long-term value of $\dot{P}=1.43(1)\times10^{-11}$~s\,s$^{-1}$ \cite{OlausenK2014, IsraelER2016}. 
This result suggests that the spin behavior of \src\ significantly changed after the initial activity detected on October 10 \cite{MereghettiGF2022}.

To ensure high accuracy in tracing the TOAs, we divided the \nicer\ observations into segments of integer orbits. 
Each segment was chosen to contain a sufficient number of photons to give a small uncertainty of $\sigma_{\phi}\lesssim0.05$ for each TOA. 
Near the peak of the flare, a \nicer\ orbit was further divided into three segments owing to a large number of collected photons. 
The TOAs were determined using the unbinned maximum-likelihood method \cite{LivingstoneRC2009, RayKP2011}. 
A pulse profile with the highest signal-to-noise ratio was selected to construct a nonparametric template by smoothing it with a Gaussian filter and interpolating it to a bin size of $10^{-4}$ cycles. 
This same step size was used to compute the phase shift $\Delta\phi$ of each TOA. 
The same analysis procedure was also performed on the \nustar\ data in the 3--8 keV energy range.

\begin{figure}[h]%
\centering
\includegraphics[width=0.9\textwidth]{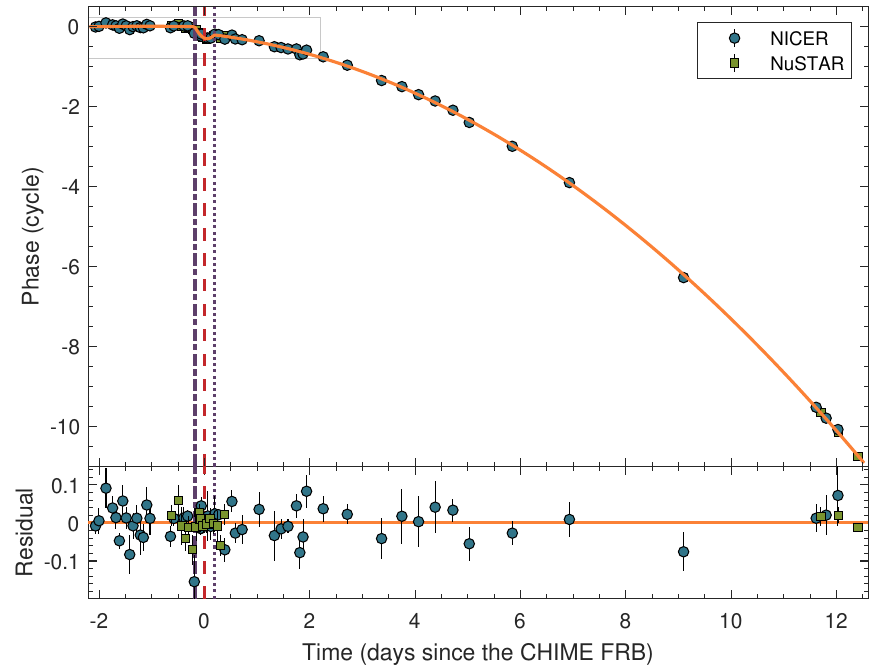}
\caption{\revi{Phase evolution of TOAs between 2022 October 12 and November 06.} The orange line is the best-fit two-glitch model, where the red vertical line denotes the time of the CHIME FRB detection. The times of two glitches are shown as the vertical dashed-dotted line (first glitch) and the dotted line (second glitch). The gray box denotes the zoom-in region shown in Figure \ref{fig:fig1_toa_analysis}. The residual is shown in the lower panel. }\label{fig:toa_analysis_supl1}
\end{figure}

Then, we traced the phase evolution of the TOAs based on the aforementioned ephemeris (see Extended Data Figure \ref{fig:toa_analysis_supl1}):
\begin{equation}
\phi(t) = \nu(t-t_0) + \frac{1}{2}\dot{\nu}(t-t_0)^2 + \frac{1}{6}\ddot{\nu}(t-t_0)^3+\ldots.
\label{eq:quadratic_model}
\end{equation}
where those terms higher than the first derivative term ($\dot{\nu}$) are ignored \revi{for the first trial}. 
It was clear that the initial timing solution failed to accurately predict the subsequent evolution. 
The phases of the TOAs showed a drop of approximately 0.2 cycles over a timescale of $\lesssim0.5$ days, starting about 4.4 hours prior to the CHIME FRB detection.
Interestingly, the evolution formed a concave quadratic curve, indicating a fast spin-down over half a day. 
After approximately 4.4 hours following the CHIME FRB, the evolution showed a long-term convex quadratic curve, implying a much slower spin-down thereafter. Consequently, we fit the TOAs with a two-glitch model:
\begin{equation}
\begin{cases}
\phi(t)\equiv\phi_0(t) = \nu(t-t_0) + \frac{1}{2}\dot{\nu}(t-t_0)^2 \textrm{~~ for~~} t<t_{g1} \\
\phi(t)\equiv \phi_1(t)=\phi_0(t) + \Delta\nu_1(t-t_{g1}) + \frac{1}{2}\Delta\dot{\nu}_1(t-t_{g1})^2 \textrm{~~ for~~} t_{g1}<t<t_{g2}\\
\phi(t)=\phi_1(t) + \Delta\nu_2(t-t_{g2}) + \frac{1}{2}\Delta\dot{\nu}_2(t-t_{g2})^2 \textrm{~~ for~~} t>t_{g2}
\end{cases}
\label{eq:glitch_model}
\end{equation}
where $t_{g1}$ and $t_{g2}$ are glitch times.
The sudden changes in frequency and spin-down rate are denoted as $\Delta\nu_1$ and $\Delta\dot{\nu}_1$ for the first glitch, and $\Delta\nu_2$ and $\Delta\dot{\nu}_2$ for the second glitch, respectively. 
The best-fit parameters for this model are listed in Extended Data Table \ref{tab:timing_solution}. 
The best-fit timing model is shown as the orange curve in Figure \ref{fig:fig1_toa_analysis} and Extended Data Figure \ref{fig:toa_analysis_supl1}. 
This model yields a fit statistic $\chi^2$ value of 83.2 with 64 degrees of freedom, \revi{and the residual did not show significant long-term trends.}

%The initial values are $t_0=$MJD$59865.6782217$ $\nu=0.3075276$ Hz, and $\dot{\nu}=-2.02\times10^{-11}$ Hz s$^{-1}$.

\begin{figure}[h]%
\centering
\includegraphics[width=0.99\textwidth]{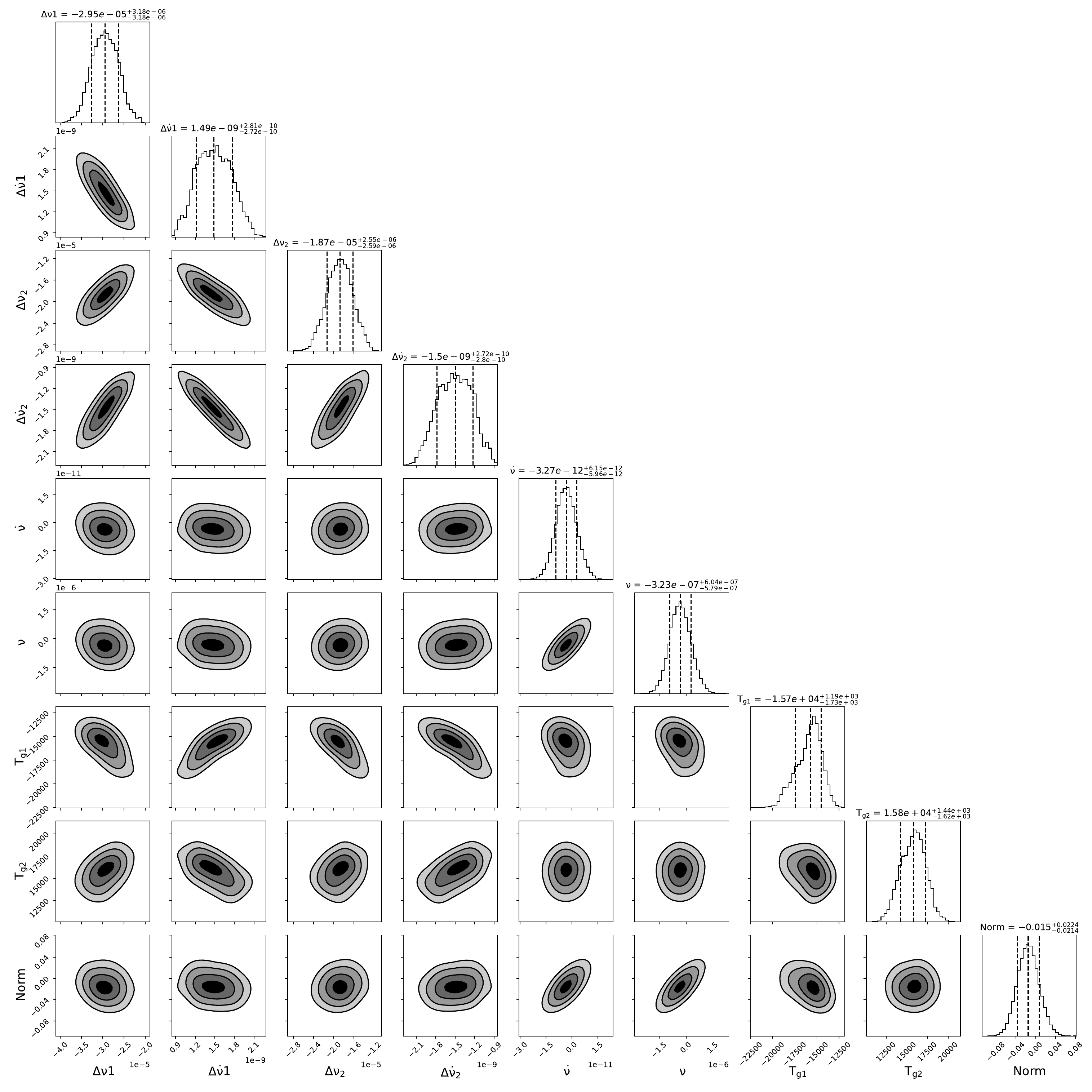}
\caption{\revi{Posterior probability density distributions of the emcee sampler} The total simulation steps is 100,000. Two spin-up glitches are needed, where the first one is accompanied by an increase in the spin-down rate, while the second glitch is accompanied by a decrease in the spin-down rate. The change in the spin-down rate of these two glitches is mostly canceled out. In the 1-D histograms, the dashed lines represent the best-fit value along with its $1\sigma$ standard deviation. }\label{fig:corner_plot}
\end{figure}

The best-fit parameters and their corresponding 1-$\sigma$ uncertainties were obtained through an MCMC simulation. 
We sampled the full parameter space of the two-glitch model using the \texttt{emcee} MCMC sampler\cite{ForemanHL2013} and assuming flat prior probability densities for all parameters. 
The MCMC was run with 32 walkers for $10^5$ steps, and the first $10^4$ samplings were discarded. 
The posterior probability density distributions are displayed in Extended Data Figure \ref{fig:corner_plot}, and the values and 1-$\sigma$ uncertainties are summarized in Extended Data Table \ref{tab:timing_solution}. The 1-$\sigma$ intervals of the best-fit model are plotted as the gray shaded area in Figure \ref{fig:fig1_toa_analysis} and Extended Data Figure \ref{fig:toa_analysis_supl1}.

\begin{figure}[h]%
\centering
\includegraphics[width=0.9\textwidth]{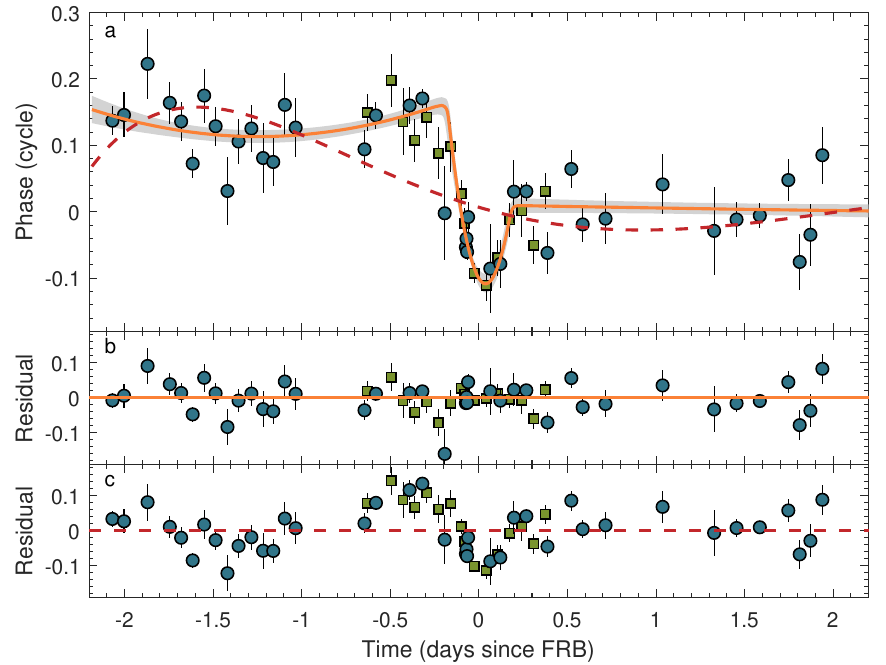}
\caption{\revi{Phase evolution of TOAs determined with \nicer\ and \nustar\ based on the post-glitch ephemeris}. The symbols used here is the same as those in Extended Data Figure \ref{fig:toa_analysis_supl1}. The red dashed curve is the best-fit polynomial ephemeris up to the 8$^{\textrm{th}}$ order time derivative of $\nu$. This model has exactly the same number of free parameters as that of the two-glitch model. The residual of the two-glitch model is shown in panel (b), while that of the 8$^{\textrm{th}}$-order polynomial is shown in panel (c).  }\label{fig:toa_analysis_supl2}
\end{figure}

In comparison to the pre-glitch epochs, the post-glitch ephemeris is more precisely defined owing to its long time span. To verify our timing solution, we cross-checked it by calculating the phases of the TOAs based on the post-glitch second-order ephemeris with a reference epoch of $t_0=$MJD 59868.87136, $\nu=0.3075264$~ Hz, and $\dot{\nu}=-4.9\times10^{-12}$~Hz\,s$^{-1}$. 
A zoom-in view of the phase evolution near the FRB is presented in Extended Data Figure \ref{fig:toa_analysis_supl2}. 
The parameters and the $\chi^2$ statistic of the best-fit model are in full agreement with those obtained previously. 
Furthermore, we tried to fit the phase evolution with an $8^{\textrm{th}}$-order polynomial, including terms up to $d^8\nu/dt^8$ in Equation \ref{eq:quadratic_model}. 
Although this high-order polynomial has the same number of free parameters as the two-glitch model, it results in a much higher $\chi^2=378.8$ with 64 degrees of freedom. 
The residuals (as shown in panel c of Extended Data Figure \ref{fig:toa_analysis_supl2}) exhibit a noticeable long-term trend, implying that the two-glitch model provides a much better fit.

The two glitches observed in this outburst are among the largest of all glitches determined in pulsars and magnetars, as demonstrated in Figure \ref{fig:glitch_size_distribution}. 
We plotted the values of $\Delta\nu$ versus $\Delta\dot{\nu}$ and compared them to previously recorded glitches in pulsars and magnetars obtained from the literature and the Jodrell Bank Glitch Catalogue \cite{EspinozaLS2011, HuN2019, YounesBH2023, GeYL2023}. 
%These two glitches are located in the upper right of the figure and follow the linear trend of the regular glitches observed in other pulsars and magnetars.

\revi{The rotational energy change involved in a glitch is $E_{\rm{glitch}}\approx 4\pi^2I\nu\Delta\nu$, where $I$ ($\approx10^{45}$~g~cm$^2$) is the stellar moment of inertia. Thus the energies associated with these two glitches are $3.9\times10^{41}$~erg and $2.6\times10^{41}$~erg. The change in rotational energy between the glitches is estimated to be
\begin{equation}
  \Delta E_{\rm{rot}}=\int_{t_{g1}}^{t_{g2}} \dot{E}_{\rm{rot}} \, dt \;\approx\; -6.5\times 10^{41} \,\, \rm{erg,}
\end{equation}
where $\dot{E}_{\rm{rot}} = 4\pi^2I\nu\dot{\nu}$ is the inferred rotational energy loss rate during the period of enhanced torque. This value of $\Delta E_{\rm{rot}}$ is approximately equal to the total energy associated with the two glitches.}

\begin{figure}[h]%
\centering
\includegraphics[width=0.9\textwidth]{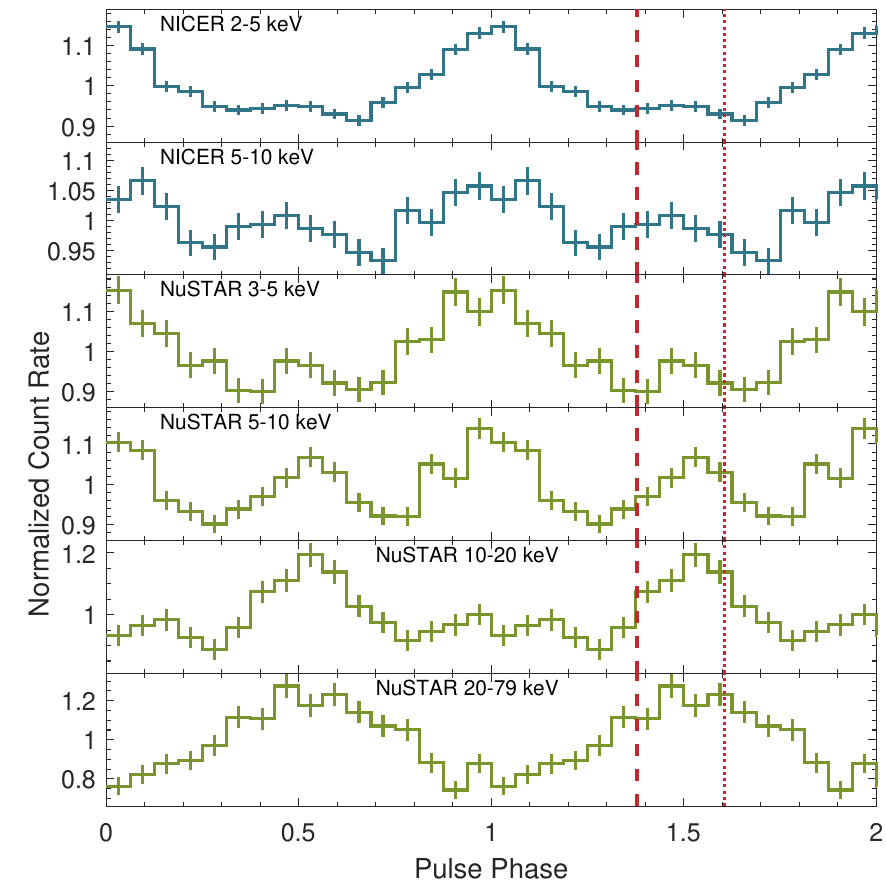}
\caption{\revi{Energy-resolved pulse profile of \src.}. Pulse profiles obtained with \nicer\ are divided into 2--5~keV, 5--10~keV bands, where \nustar\ data were divided into 3--5~keV, 5--10~keV, 10--20~keV, and 20--79~keV bands. The red dashed line and dotted line indicate the phase of the FRB emission detected by CHIME and GBT,  respectively.}\label{fig:profile_diff_energy}
\end{figure}

\noindent \textbf{Energy-Resolved Pulse Profile:} 
We investigated the energy dependence of the pulse profile to gain insights into the emission mechanisms. 
To this end, we obtained the energy-resolved pulse profiles based on the best-fit two-glitch model by dividing the \nicer\ data into two energy bands, 2--5~keV and 5--10~keV, and the \nustar\ data into four energy bands, 3--5~keV, 5--10~keV, 10--20~keV, and 20--79~keV. 
The results are presented in Extended Data Figure \ref{fig:profile_diff_energy}. 
The soft X-ray peak is evident at a phase of $\phi\approx0$, while the hard X-ray peak is found at $\phi\approx0.5$. 
The soft X-ray peak dominates the profile below 10 keV but vanishes at energies above 20 keV. 
On the other hand, the hard X-ray peak is present above 5 keV and becomes the dominant feature above 10 keV.

We also calculated the spin phase of the FRB. 
The barycentric corrected time of the CHIME FRB\cite{Dong_CHIME_2022} was calculated to be MJD 59866.80817034 using \texttt{pintbary}, a command line tool of the pulsar timing software package \texttt{pint}\cite{LuoRD2021}\footnote{\url{http://ascl.net/1902.007}}. Similarly, we calculated the barycentric correct time of the brightest of the GPT FRB\cite{MaanLS2022} as MJD~59866.80817889. \revi{These times were extrapolated to infinite frequency based on a dispersion measure of 332.8~\cite{Dong_CHIME_2022}.}
The phase of the FRB was found to be approximately 0.38 based on CHIME detection or 0.5 based on GBT result, which is close to the peak of the 20-79 keV band (see Extended Data Figure \ref{fig:profile_diff_energy}). 
%This suggests that the FRB emission is likely more related to the magnetospheric activity rather than hotspots on the surface.

\noindent \textbf{Verification of the spin-down rate between glitches:} 
From the analysis of the energy-resolved pulse profile, it is evident that the pulsed emission has multiple origins. 
The soft X-ray band may be dominated by thermal emission from the neutron star's surface, whereas the hard X-ray emission would be originating from the magnetosphere. 
This suggests that the timing solution based on photons below 8 keV could be affected by the pulse profile change due to the motion of hotspots on the surface during the outburst. 
To investigate this, we attempted to trace the evolution of the phase of the 20-79 keV profile, but the large uncertainty due to the limited number of photons collected (approximately 2200 counts between the two glitches) made it difficult to trace the evolution in detail.

\begin{figure}
\includegraphics[width=0.99\textwidth]{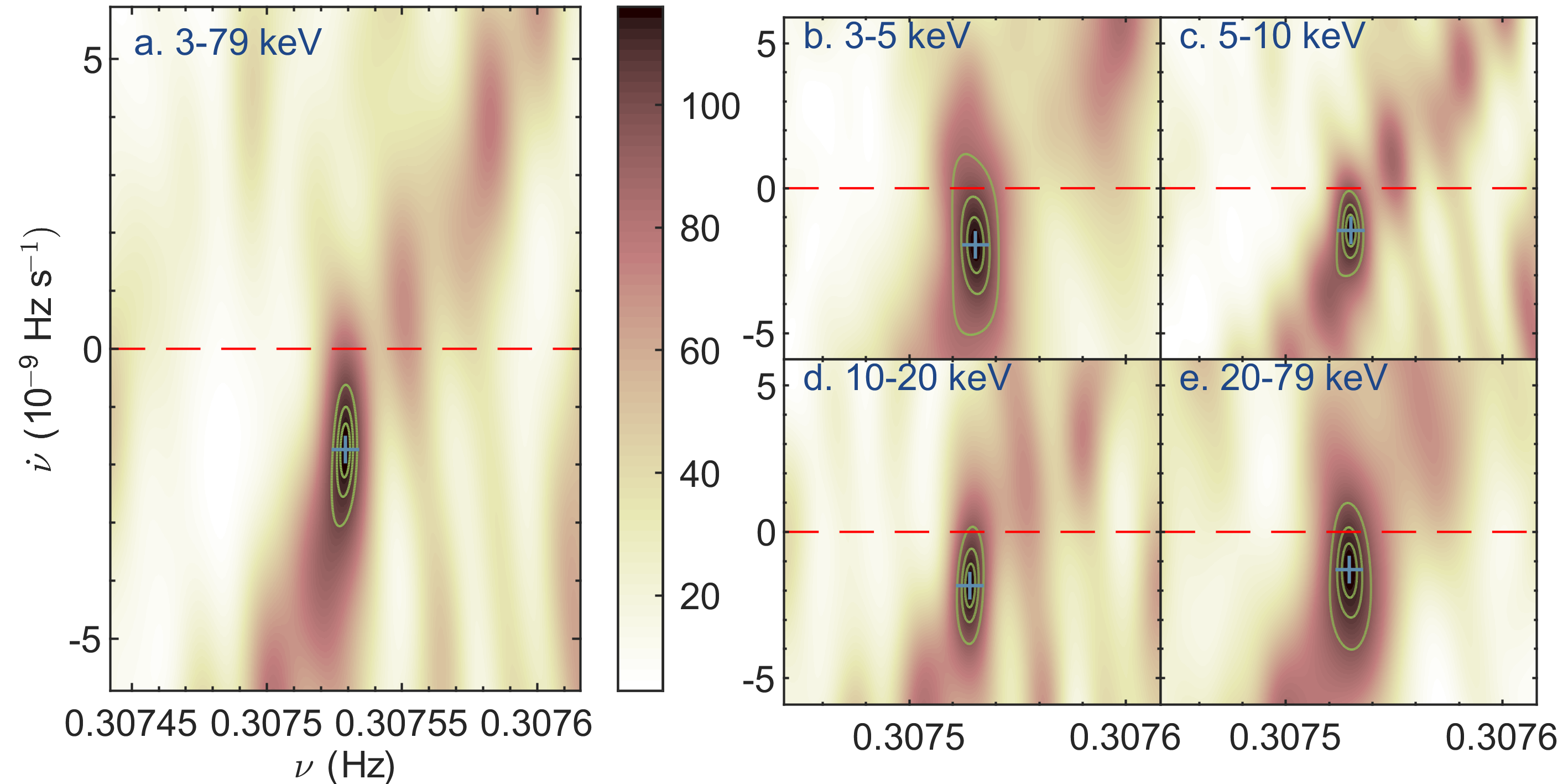}
\caption{\revi{Two-dimensional $Z_2^2$-test searching result between two glitches} We performed search in the a.~3--79 keV, b.~3--5 keV, c.~5--10 keV, d.~10--20 keV, and e.~20--79 keV bands using \nustar\ data.  X- and y-axes of panels b--d are the same as that of panel a. The color map of each panel denotes the $Z_2^2$ values, while the peak is marked with the green plus sign. The green contours denote $Z_2^2(\mbox{max})-2.3$, $Z_2^2(\mbox{max})-6$, and $Z_2^2(\mbox{max})-15$ where $Z_2^2(\mbox{max})$ is the maximum $Z_2^2$ value of the peak. \label{fig:z2test_diff_energy}}
\end{figure}

As an independent verification, we performed an energy-resolved two-dimensional Z$^2_2$-test search for $\nu$ and $\dot{\nu}$ between the two glitches. 
Using \nustar\ events collected between the two glitches, we divided the events into four energy bands as defined in the energy-resolved pulse profile analysis. 
The number of photons in each energy band was 7991 in the 3--5 keV range, 14478 in the 5--10 keV range, 6484 in the 10--20 keV range, and 2275 in the 20--79 keV range.
The results of the two-dimensional Z$^2_2$-test for each energy band, as well as the broadband (3--79 keV) events, are presented in Extended Data Figure \ref{fig:z2test_diff_energy}. 
The result obtained from the broadband events suggests a spin-down rate of $\dot{\nu}=-1.7(4)\times10^{-9}$~Hz~s$^{-1}$, which is in full agreement with the value obtained from the TOA analysis [$-1.5(3)\times10^{-9}$~Hz~s$^{-1}$]. 
The spin-down rate in each energy band was determined to be $-1.9(9)\times10^{-9}$~Hz~s$^{-1}$ for 3--5~keV, $-1.5(5)\times10^{-9}$~Hz~s$^{-1}$ for 5--10~keV, $-1.8(7)\times10^{-9}$~Hz~s$^{-1}$ for 10--20~keV, and $-1.3(9)\times10^{-9}$~Hz~s$^{-1}$ for 20--79~keV. 
These results point to a large spin-down rate that is consistent with that obtained from the TOA analysis. 

\revi{Moreover, we obtained the soft X-ray profile 2--8 keV \nicer\ data and 10--79 keV \nustar\ data in three epochs: $t<t_{g1}$, $t_{g1}<t<t_{g2}$, and $t>t_{g2}$. Those X-ray photons collected one hour close to the glitch epochs are not included to avoid possible coverage of the glitch time. No significant phase shift can be observed. These suggest the recovered phase evolution is secure and not a misinterpretation of hotspot migration.}

\noindent \textbf{Spectral Analysis:}  %\textcolor{red}{(Takuto and Teru)}
%The \nicer\ spectral data were reprocessed with NICERDAS version 9 as part of the \texttt{HEASOFT} version 6.30.1 using the calibration database version 20210707. 
%The unfiltered event data were calibrated and screened with the \texttt{nicerl2} pipeline by using default parameter settings as the standard \nicer\ analysis.
%The unfiltered event data were calibrated and screened with the \texttt{nicerl2} pipeline by using default parameter settings, producing scientifically available cleaned event lists.
%The \nustar\ spectral data were extracted with the analysis software \texttt{NUSTARDAS} version 2.1.2 as a part of \texttt{HEASOFT} version 6.30.1 and the calibration database of version 20220926. We performed data calibration and screening using the \textit{nupipeline} script and extracted the spectra with \textit{nuproduct}.
%The background region of \nustar\ data was selected from a source-free region with a circle of a 2.5 arcmin radius. 
We extracted \nicer\ and \nustar\ (FPMA and FPMB) persistent spectra with burst-removed and barycentric corrected GTIs (Extended Data Table~\ref{tab:observation_infomation}).
\nicer~spectra were extracted with \texttt{xselect} version 2.5a following the standard procedure, where the background of \nicer~data was produced with \nicer~background estimator tool \textit{nibackgen3C50} version v7b \cite{2022AJ....163..130R}.
The persistent spectra of epochs A, D, and E are extracted from the time intervals where \nicer\ and \nustar\ observed the target simultaneously, whereas those of epochs B, C, and F do not strictly rely on time intervals with simultaneous observations.
%The persistent spectra of the quiescent epoch and epoch F are extracted with all GTIs of each observation without removing bursts since less significant bursts or flares were detected.
We also extracted the burst spectra of \nicer\ and \nustar\ (FPMA and FPMB) by accumulating all the detected burst at epochs B, C, and D using the barycentric corrected burst GTIs. 

\begin{figure}
\includegraphics[width=1.0\textwidth]{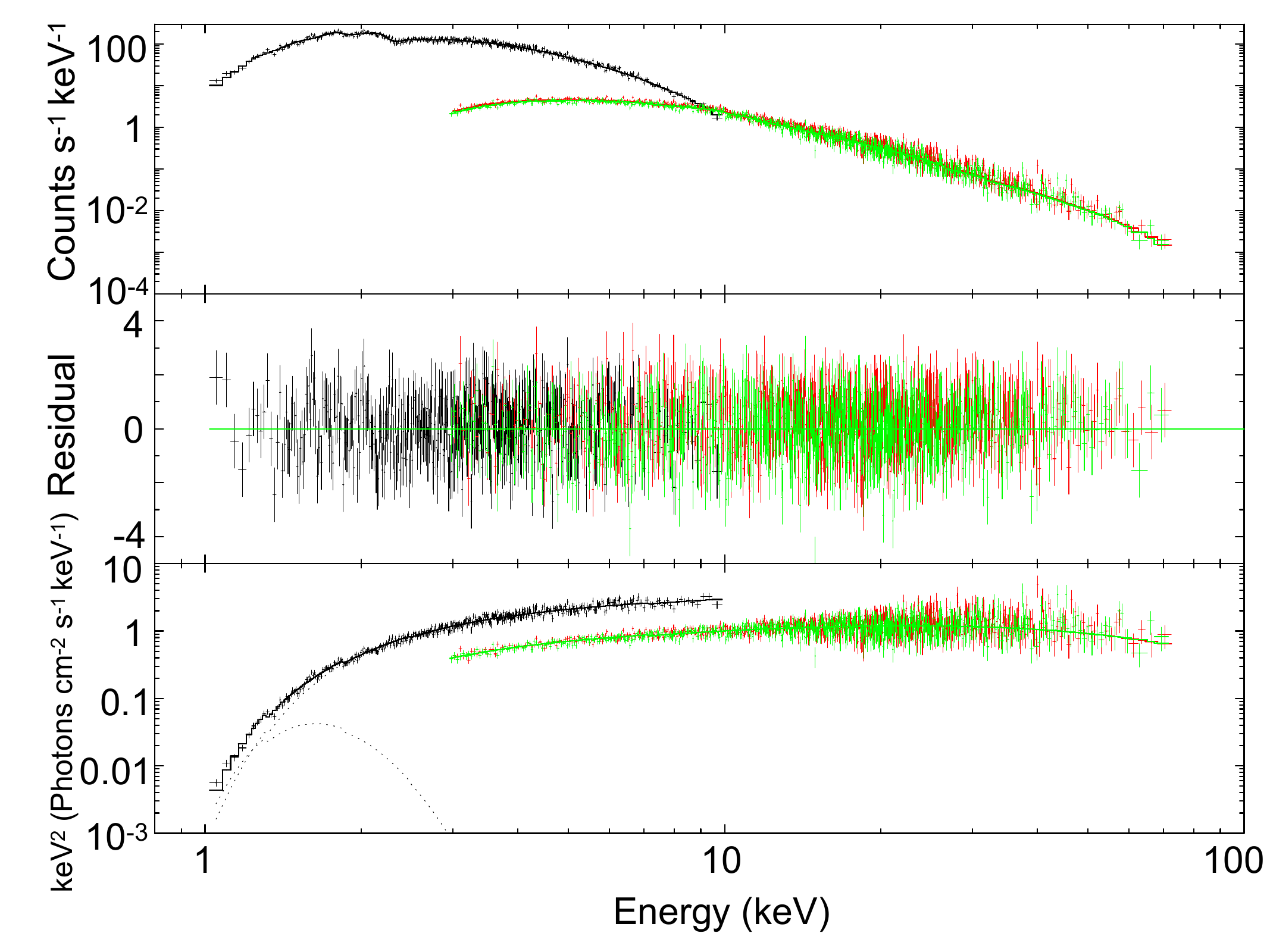}
\caption{\revi{Spectral fitting results for burst emission from \src\ at the epochs between two glitches (Epochs B, C, and D).}
The data are extracted from \nicer~(black), \nustar~FPMA(red), and \nustar~FPMB (green). All the detected burst events are accumulated. The total exposure of the NuSTAR data is 824~s. The accumulated spectrum is fitted by a model with an absorbed blackbody plus a power law with an exponential roll-off. Top, middle, and bottom panels show the count spectral components, fitting residuals, and spectra in $\nu F_{\nu}$ form, respectively.  \label{fig:spectral_fit_results_burst}}
\end{figure}

%% note
We used the \texttt{XSPEC} software version 12.12.1 with the Chi-square statistic in the following analysis.
We fitted the persistent spectra at each epoch (Figure \ref{fig:fig_spectrum}) and the burst accumulated spectrum (Extended Data Figure \ref{fig:spectral_fit_results_burst}), where the information of observation and the best fit spectral parameters are shown in Extended Data Table \ref{tab:best_fit_spectral_parameter} and \ref{tab:observation_infomation}, respectively.
The absorbed blackbody (BB) and power law (PL) model, \verb|(tbabs*(bbodyrad+pegpwrlw))|, was used for the persistent spectra except for epoch C.
The solar abundance value was used to calculate the absorption \citep{Wilms2000}.
An absorbed PL model \verb|(tbabs*pegpwrlw)| was used for epoch C since the BB component is negligible compared to the PL.
The accumulated burst spectrum was fitted with an absorbed BB plus a PL with a high energy exponential roll-off \verb|(tbabs*(bbodyrad*cutoffpl))|, hereafter CPL.
We added a constant term ($\sim 3$) in the spectral model to cross-calibrate the normalization between \nicer\ and \nustar\ because these two instruments cover emissions from different bursts.
Best-fit spectral parameters, including the hydrogen column density $N_{\rm H}$, BB temperature $kT$, its radius $R$, and the photon index $\Gamma$ of the PL, are tabulated in Extended Data Table \ref{tab:best_fit_spectral_parameter}. These parameters were set free except for the spectra of epoch B, where the BB parameters are frozen to those of epoch A since no \nicer\ observation was available and \nustar\ spectral shape was similar to that of epoch A than that of epoch C.

The absorbed 2--70 keV X-ray flux of persistent spectra was calculated with the \verb|flux| command of \texttt{XSPEC} and the \nicer\ and \nustar\ spectra were used for low (2--10 keV) and high (10--70 keV) energy bands, respectively.
The absorption-corrected flux of persistent spectra was calculated from the normalization factors of the BB and PL models.
The absorbed and absorption corrected flux of burst spectra was calculated from normalization factors of the BB and PL models with only \nustar\ spectra because of the difficulty of estimating the effect of the constant value.
%The flux of the black body was calculated with \verb|cflux|, which is the model for calculating flux on Xspec and in which the norm of black body was set to 1 and the other parameters were fixed.
%The flux of the power law was caluclated with \verb|pegpwrlw|, including the flux in the model parameter. 
%In both spectral models, we set the energy bands to 2--70 keV for our flux calculations.

The hydrogen column density $N_{\rm H}$ was determined to be $(2.1$--$3.4)\times10^{22}$~cm$^{-2}$ from the persistent spectra.
Comparing the quiescent spectra, the BB temperature $kT$ increased to 0.51--0.68 keV around the FRB (epoch A--E) and returned to the quiescent value 12 days after the FRB (epoch F).
On the other hand, the BB radius $R$ was similar to the quiescent value. 
Around the peak of the count rate in X-ray bands (Epoch C), the power-law spectrum became softer (from $\Gamma \sim1.2$ to $1.6$) and extended from the hard X-rays to lower energies, making it difficult to measure the BB parameters.
%Before and after the FRB, the photo index of the power law spectra increased from $\sim1.2$ to $\sim1.7$.
In contrast, burst spectra have a high hydrogen column density ($4.2^{+0.2}_{-0.1}\times10^{22}~\rm{cm^{-2}}$) and a soft photon index ($1.40\pm0.02$). 
%Burst spectra has high hydrogen column density ($4.15^{+0.4}_{-0.3}\times10^{22}~\rm{cm^{-2}}$) and the soft photon index ($1.702\pm0.006$). 

We also compared the above fitting results around the glitch (epochs A -- E) with those considering \textsl{SCORPEON} background model (obtained with the tool \texttt{nicerl3-spec} of NICERDAS v10).
A few fitting parameters vary from the values obtained with the 3C50 background model, but the difference is at most around 20\% (see 
Extended Data Table~\ref{tab:best_fit_spectral_parameter}). The difference could be larger than their 1$\sigma$ statistical uncertainty and could be inherent in modeling the background. However, they did not change our order-of-magnitude estimate of the total energy and the spectral evolution. 

\noindent \textbf{Theoretical Interpretation:} 
The main deliverables of this unprecedented observational campaign are the two spin-up glitches that bracket a strong spin-down epoch that persists for several hours, the enhanced soft and hard X-ray activity evident mostly between the glitches, and the emission of the FRB during this activity interval.  All these arise during an active epoch of X-ray bursts.  The ephemeral wind picture presented here to address these features is derived from that presented in \cite{YounesBH2023} for the spin-down glitch observed from \src\ in 2020 during a period where source monitoring was much sparser than for this event. 
%\revi{We note that there are other possible interpretations. A rapid spin-down epoch could arise if there is a corresponding increase in the oblateness and/or moment of inertia $I$ \cite{Mastrano-2015-MNRAS}.  Alternatively, angular momentum transfer to a more slowly spinning inner crust could seed such a timing signature \cite{Thompson-2000-ApJ}.  Detailed exploration of these models in the context of the double glitch is the purview of future work.} 

\noindent \textbf{Constraining the Superfluid Component:}
\revi{First, the focus is on the timing results.
Spin-up glitches impart abrupt re-adjustments to the rotational equilibrium of the entire star~\cite{Anderson1975,HaskellM2015}.  
Large glitch sizes typically require moments of inertia $I_{\rm sf}$ of the superfluid neutron component in the inner crust of around a few percent of that ($I$) of the entire star \cite{Link1999,Andersson2012, Chamel2013, Ho2015}.  This ratio can be estimated using angular momentum conservation arguments \cite{alparetal81,Link1999}, % for the entire double glitch episode,
yielding
\begin{equation}
    {{I_{\rm sf} \over {I}}} \;\sim\;
    {{\Delta\nu} \over {\lvert\dot{\nu}\rvert \Delta T_{\rm g}}}
 \label{eq:sf_fraction}
\end{equation}
where $\Delta T_{\rm g}$ %$=t_{g2}-t_{g1}$
is the time interval between glitches and is $\approx 8.8$~hours in the case here.  This estimate suggests a superfluid component comprising several tens of percent of the entire star.}

\noindent \textbf{The Inter-Glitch Torque driven by a  Strong Wind:}  
%
%Historically, conventional spin-up glitches for pulsars are explained in terms of the sudden redistribution of angular momentum inside the neutron star to address the differential between a fast-spinning interior and a slower-spinning, magnetized crust/outer crust; this is addressed in the main text above.
The rapid spin-down episode following the first glitch leads to a net increment $\Delta\nu \sim -4.7 \times 10^{-5}\,$Hz in the hours prior to the second glitch.  This evolution could be attributed to an external torque on the star and likely linked to the magnetosphere.  The October 2020 anti-glitch ($\Delta\nu=-1.8\times10^{-6}$ Hz) detailed in \cite{YounesBH2023} was interpreted as due to a strong, ephemeral wind lasting around 10 hours or less in its peak phase.  This wind loads portions of the magnetosphere that are tied to the active surface region with plasma, altering the magnetic configuration and opening field lines near the pole \cite{Mahlmann2022ApJ}.  Thus the wind introduces a monopolar component to the field geometry that renders the polar fields more radial, approaching that of the morphology and strength of a split monopole \citep{1995ApJ...443..810W,TLK-2002-ApJ,Hu-2022-ApJ} in the limit of extremely large winds and strong twists.  

Such a picture can be also envisaged for the \src\ timing data for October 2022 presented here.  The expected crustal stresses associated with the first glitch, amplified by the superstrong magnetic field, could precipitate plastic flows or rupturing, and localized heating and mechanical transport that could eject plasma from the surface near the magnetic pole into the magnetosphere. There, the locally zenith-directed fields permit rapid energy and mass transport vertically, and the thermal radiation pressure is above the Eddington limit, driving baryon-loaded plasma into the magnetosphere as a wind. Only mass transfer on open field lines can shed angular momentum to infinity and slows down the crust's rotation.  

The wind required to drive the high spin-down episode between the glitches is highly optically thick \cite{YounesBH2023}, yet quite confined in its magnetic colatitudes in the inner magnetosphere.  The polar cap colatitude of the last open field line footpoint at the surface for a dipolar configuration is $\theta_{\rm cap} \approx \sqrt{2 \pi R_{\rm ns}/Pc} \approx 0.46^{\circ}$ for a rotational period of $P=3.25$~sec.  Thus, even with a modest expansion of the polar cap due to twists at the surface, the wind collimation at low altitudes is sufficient that it does not significantly obscure the surface soft X-ray emission: the inferred blackbody emission zone radii of $R_{\rm bb}\sim 2$\,km in Extended Date Table 4 suggest that the soft X-ray hot spot subtends angles $\theta_{\rm bb}\gtrsim R_{\rm bb}/(R_{\rm ns}\sqrt{\pi}) \sim 7^{\circ}$.  This is for a source distance of 6.6 kpc, which is not well constrained, \revi{thereby introducing some uncertainty into this} $\theta_{\rm bb}$ estimate.  \revi{Near the surface, the wind is far less dense than the plasma that radiates the much more impulsive magnetar giant flares, and its motion is much less relativistic \cite{YounesBH2023}, thereby not being subject to strong Doppler boosting.  Accordingly, its effective temperature lies at or below the \nicer\ band \cite{YounesBH2023}, indicating that the wind's periphery may contribute to the soft X-ray signal, but not overwhelm it with hard X rays.}  Yet such a wind does not occult the persistent hard X-ray signals at relatively low altitudes on closed field lines (see below).  

The analysis of ephemeral mass loss delivered in \cite{YounesBH2023} that leads to the conclusion of high optical depth in the wind needs to be adapted for the high $\vert\Delta\nu\vert$ observed here between the glitches. Using total angular momentum conservation, the time-integrated mass loss $\delta m$ for a star of mass $M$ and radius $R_{\rm ns}$ can be expressed via the ratio $\delta m/M \sim 2/5 \, (R_{\rm ns}/R_{\rm eq})^2\, \vert \Delta \nu \vert /\nu$ for a spin down increment $\Delta\nu$ in $\nu$; see Eq.~(3) in \cite{YounesBH2023} and associated discussion for details.  Here $R_{\rm eq}$ is the radius of equipartition between the energy densities of the flowing plasma and the magnetic field.  For the dipole field geometry focused on in \cite{YounesBH2023}, one quickly infers $R_{\rm eq} \sim 5 R_{\rm ns}$ and a mass loss ratio $\sim 2.4 \times 10^{-6}$.  This high value is not a realistic estimate because the plasma loading of the polar magnetosphere is sufficient to enhance the local field strengths \citep{Hu-2022-ApJ} substantially when approaching a split monopole configuration, wherein the field strength scales roughly as $1/R$ with altitude $R$. This twisted field enhancement moves the equipartition radius to around $R_{\rm eq} \sim 20-50 R_{\rm ns}$, leading to the estimate of a fraction $\delta m/M\sim 2.4 \times 10^{-8} - 1.5 \times 10^{-7}$ of the stellar mass expelled during a nominal 8.8-hour wind phase. This constitutes a mass loss roughly on the scale of that expected for giant flares. To provide a more precise estimate requires a detailed study of the dynamics of radiation-coupled plasma winds and angular momentum expulsion in magnetar magnetospheres; this will be the subject of a future study.

This wind can persist as long as the heat/mechanical energy source near the crustal active zone is maintained, and then abates as the surface regions ``heal,'' and the polar magnetosphere untwists. Observationally this ``volcanism'' timescale is $\lesssim 9$ hours, and this provides an important constraint for theoretical models for how sub-surface conditions associated with magnetar glitches are generated.  The cumulative spin-down during this epoch was sufficient to return the rotational configuration in $\sim 9$ hours to a critical state; this may be why a second glitch of similar magnitude ensued.   The wind's abatement suggests at least partial crustal stress relaxation or healing.  Note that mass that is shed on peripheral closed field lines (spillover ``wind'') can potentially influence emission from the closed portions of the atmosphere and magnetosphere.  

\noindent \textbf{Enhanced X-ray Activity between the Glitches:}
Two hours after the first glitch, there is a sharp rise in the burst rate and the persistent hard X-ray emission flux from \src\ (see Figures~\ref{fig:fig1_toa_analysis} and \ref{fig:fig_spectrum}).  Assuming that these radiative changes are associated with the glitch, the time delay can be used to constrain the crustal location of heating within the magnetar, via the thermal timescale $\tau_{\rm th}\sim (C/\kappa)z^2$, where $C$ is heat capacity, $\kappa$ is thermal conductivity, and $z$ is depth below the surface.  For typical $C$ and $\kappa$ of strongly magnetic neutron stars (see, e.g., \cite{Potekhin1999}), a two-hour thermal timescale occurs at a depth $z\sim 10\mbox{ m}$ and density of $10^7$~g cm$^{-3}$ for polar locales where efficient heat transport occurs along {\bf B} approximately in the radial direction. This timescale likely exceeds that for the onset of the wind, since this outflow emanates from very near the surface; yet it may indicate when the wind enters its most powerful phase.  The temporal correlation of the burst rate and persistent emission enhancement and decline following the first glitch suggests they are driven by the same sub-surface energy deposition, with concurrent evolution and relaxation of the crust, atmosphere, and magnetosphere. 
%\mgbc{\sout{The decline of the X-ray burst rate on hours timescale is reminiscent of the temporally log-uniform rate of radio bursts exhibited by some cosmological FRBs on similar timescales \citep{WadiasinghT2019,2020ApJ...902L..17T} .}}

Concerning the spectroscopy, there may be a causal coupling between the observed heating of the surface soft X-rays in the \nicer\ data and the enhanced persistent signal seen above 10 keV by \nustar\ between the glitches.  The leading paradigm for the generation of the persistent hard X-ray emission above 10 keV in magnetars is via resonant inverse Compton scattering (RICS) of soft X-ray photons emanating from the stellar surface by relativistic electrons accelerated in the magnetosphere \citep{BH-2007-ApSS,FT-2007-ApJ}.  The scattering cross section is resonant at the cyclotron frequency.  The electrons are accelerated by magnetospheric electric fields, coherent or non-stationary, and given that the \nustar\ signal is enhanced and not abated between the glitches, we can presume that the collimated wind does not impair lepton acceleration in the pertinent electric field zone. This region is likely on closed field lines as most of the RICS signal, which is highly anisotropic and beamed, originates in quasi-equatorial locations where the magnetic field is low enough for the scattering to access the cyclotron resonance.  Moreover, for 
these locations, an observer who samples the Doppler-boosted RICS emission along field tangents \citep{WBGH-2018-ApJ} will naturally see a hard X-ray pulse peak at a different phase from that for the soft X-ray surface signal, as is evident in Extended Data Figure 4.  The intensity of the RICS emission is roughly proportional to the number density of the surface thermal photons \citep{BWG-2011-ApJ}, which for a blackbody scales as $T^3$.  Accordingly, the roughly 60\% increase in $T$ between the quiescent epoch and interval D as listed in Extended Data Table 4 should naively yield an enhancement in the RICS signal by a factor of $\sim 4$.  In addition, the temperature increase might significantly extend the volumes of the inner magnetosphere for which the cyclotron resonance that controls the RICS emissivity is accessible, thereby further increasing the RICS flux.
Combined, these influences would account for about half or more of the enhancement in the \nustar\ spectrum evident in Figure 3, the remaining portion perhaps due to increases in the relativistic electron density associated with a spillover ``wind.'' 

\noindent \textbf{The Fast Radio Burst:}
If the electric fields that accelerate the electrons that generate the RICS signal are left unmodified by the polar wind, then their persistence may continue pair creation that contributes to an enhanced RICS signal that can then further lead to enhanced pair cascading. The absence or presence of plasma where a burst occurs is likely connected to the generation of an FRB \cite{WadiasinghT2019, Wadiasingh2020}. Thus, the timing of the FRB during the epoch of amplified and abating hard X-ray emission between the glitches may in fact be a causal association. As a putative spillover wind abated and the system relaxed after the local persistent flux peak 2-3 hours after the first glitch (time interval C), this region became increasingly charge-starved to errant bursts during the span where the highest torque, and likely also stresses, were inflicted on the crust. This phenomenology is reminiscent of the April 2020 FRB, which also occurred in the waning phase of the burst storm on its track to relative source dormancy.  Furthermore, it is noteworthy that the phase-connected arrival time of the FRB is roughly coincident with the peak of the hard X-ray signal in Extended Data Figure~4, and offset in phase with the peak of the soft X-rays.  Thus it appears probable that in this case both the FRB and the persistent hard X-rays both originate from quasi-equatorial locations, though likely not at the same altitudes.

\noindent \textbf{The October 2022/October 2020 Glitch Comparison:}
It is of interest to compare this October 2022 event with the anti-glitch in \src\ of the October 2020 epoch, which was not afforded the same high-quality observational cadence \cite{YounesBH2023}.  The 2020 anti-glitch was inferred using timing data spanning over 50 days, and arose during a data gap of approximately 3 days duration. An \revi{FRB-like} burst was detected at about 3 days after the putative time of the anti-glitch, and pulsed radio emission ensued thereafter, both during a period when the X-ray timing was re-established and at the post-glitch ${\dot \nu}$ \cite{YounesBH2023}.  No X-ray enhancement before or after the anti-glitch was seen.  For the October 2022 events studied here, if instead, the source had been subject to a similar observational data gap of duration 1--2 days that bracketed the two glitches, none of the enhancement in persistent X-ray emission, the FRB and the glitches themselves would have been detected. In fact, the subsequent relatively sparse timing data might actually accommodate an anti-glitch solution for the October 2022 event, albeit likely of different magnitude in $\Delta \nu/\nu$.  Clearly, the two events are different, particularly in terms of the eventual pulsed radio emission seen in 2020.  Yet they do present similarities, principally rapid spin-down episodes, and this motivates the position that an ephemeral wind can drive the strong torque on the star for several hours between the two glitches. Moreover, the events do pose the question of what sub-surface physics dictates a glitch recurrence timescale of around $\tau_{\rm rec}\sim 2$ years in \src. Glitch models of pulsars predict a range of $\tau_{\rm rec}$ times and distributions \citep{HaskellM2015}.  Comparing these two events clearly indicates that the excellent data cadence we present here for October 2022 highlights the rich nature of information afforded by such temporally dense observations.

%\noindent {\bf References for Methods} 

%\setcounter{enumctr}{30}
%\setcounter{#1}{30}
\let\oldthebibliography=\thebibliography
\let\oldendthebibliography=\endthebibliography
\renewenvironment{thebibliography}[1]{
    \oldthebibliography{#1}
    \setcounter{enumiv}{30}                        % Start reference from [31]
}{\oldendthebibliography}

\clearpage

\bmhead{Acknowledgments}

This work was supported by the National Aeronautics and Space Administration (NASA) through the \nicer\ mission and the Astrophysics Explorers Program. 
This research has also made use of data obtained with \emph{NuSTAR}, a project led by Caltech, funded by NASA and managed by NASA/JPL, and has utilized the NUSTARDAS software package, jointly developed by the ASDC (Italy) and Caltech (USA). 
This research has made use of data and software provided by the High Energy Astrophysics Science Archive Research Center (HEASARC), which is a service of the Astrophysics Science Division at NASA/GSFC and the High Energy Astrophysics Division of the Smithsonian Astrophysical Observatory. 
C.-P.H. acknowledges support from the National Science and Technology Council in Taiwan through grant 109-2112-M-018-009-MY3 and 112-2112-M-018-004-MY3.
T.E acknowledges RIKEN Hakubi project, JST grant number JPMJFR202O (Sohatsu), and JSPS/MEXT KAKENHI grant number 22H01267.
Z.W. acknowledges support by NASA under award number 80GSFC21M0002.
W.C.G.H. acknowledges support through grant 80NSSC22K0397 and 80NSSC23K0078 from NASA. 
M.G.B. acknowledges the support of the National Science Foundation through grant AST-1813649 and NASA through grant 80NSSC20K1564.
S.G. acknowledges the support from the CNES.
K.R acknowledges support from the Vici research programme ``ARGO'' with project number 639.043.815, financed by the Dutch Research Council (NWO).
 NICER research at NRL is supported by NASA.

\bmhead{Authors' contributions}
C.-P.H. led the data analysis, performed the timing analysis, and contributed to writing the paper. T.N., T.E., T.G., and S.G. performed the spectral analysis and contributed to writing the paper. T.E. led the \nicer\ and \nustar\ collaboration.  G.Y.~and T.E. triggered the \nicer\ DDT and joint \nicer/\nustar\ GO ToO program (\nicer\ Cycle 4 proposal number 5076), respectively. G.Y. and P.S.R. supported the timing analysis and contributed to writing the paper.  Z.W., W.C.G.H., and M.G.B. led the theoretical interpretations and contributed to writing the paper. K.R., C.K., Z.A., A.K.H, and K.C.G. contributed to writing the paper.

\bmhead{Data availability}
Availability of data and materials: NICER raw data and calibrated level-2 data files were generated at the Goddard Space Flight Center large-scale facility. These data files are publicly available and can be found \revi{here} \href{https://heasarc.gsfc.nasa.gov/FTP/nicer/data/obs/}{at this link}. \nustar\ data files are also publicly available at \href{https://heasarc.gsfc.nasa.gov/W3Browse/all/numaster.html}{NUMASTER} table. 

\bmhead{Code availability}
Reduction and analysis of the data were conducted using publicly available codes  provided by the High Energy Astrophysics Science Archive Research Center (HEASARC), which is a service of the Astrophysics Science Division at NASA/GSFC and the High Energy Astrophysics Division of the Smithsonian Astrophysical Observatory. For \nicer\ and \nustar, we used NICERDAS version v009 and NUSTARDAS version v2.1.2, respectively, part  of  HEASOFT  6.31 (\url{https://heasarc.gsfc.nasa.gov/docs/software/lheasoft}). Spectral analysis was conducted using Xspec version 12.13.0 (\url{https://heasarc.gsfc.nasa.gov/xanadu/xspec/}). The emcee MCMC sampler is a public software available at \url{https://emcee.readthedocs.io/en/stable/}. Custom codes for the timing analysis routines are available upon reasonable request from the corresponding authors.

\bmhead{Conflict of interest}
The authors declare that they have no conflict of interest.

\clearpage

%%%%%%%%%%%%%%% Extended Data Table 1: datasets %%%%%%%%%%%%%%%%%%%%%%%%
\begin{table}
\centering
    \caption{Data sets used in this work.}
    \label{tab:observation_log}
    \begin{tabular}{cccc}
    \hline
    \hline
    Observatory & ObsID & Start Date & Exposure (s) \\
    \hline
    \nicer\ & 3020560159 & 2020-10-16T21:40:00 & 1349\\
%    \nicer\ & 4020560109 & 2021-09-19T01:08:20 & 9739\\
    \nicer\ & 5020560106 & 2022-10-12T17:40:27 & 1260\\
    \nicer\ & 5020560107 & 2022-10-13T00:03:40 & 14913 \\
    \nicer\ & 5576010101 & 2022-10-13T16:44:01 & 3302 \\
    \nicer\ & 5576010102 & 2022-10-14T03:34:37 & 10332 \\
    \nicer\ & 5576010103 & 2022-10-14T23:58:20 & 7757 \\
    \nicer\ & 5576010104 & 2022-10-16T02:16:40 & 9990 \\
    \nicer\ & 5576010105 & 2022-10-17T01:14:37 & 2633 \\
    \nicer\ & 5576010106 & 2022-10-18T03:54:40 & 2472 \\
    \nicer\ & 5576010107 & 2022-10-19T04:20:01 & 2845 \\
    \nicer\ & 5576010108 & 2022-10-20T06:58:40 & 1693 \\
    \nicer\ & 5576010109 & 2022-10-21T00:00:00 & 3091 \\
    \nicer\ & 5576010110 & 2022-10-22T02:02:12 & 2034 \\
    \nicer\ & 5020560108 & 2022-10-23T21:44:37 & 260 \\
    \nicer\ & 5020560109 & 2022-10-25T06:17:53 & 678 \\
    \nicer\ & 5576010111 & 2022-10-26T08:20:34 & 6592 \\
    \nicer\ & 5576010112 & 2022-11-03T13:22:00 & 1297 \\
    \nicer\ & 5576010113 & 2022-11-06T08:16:30 & 1094 \\
    \nicer\ & 5020560110 & 2022-11-05T09:05:40 & 1194 \\
    \nicer\ & 5020560111 & 2022-11-06T00:42:00 & 729 \\
    \hline
    \nustar\ & 90602332004 & 2020-10-16T21:16:09 & 18585 \\
%    \nustar\ & 90701329002 & 2021-09-19T00:41:09 & 43696 \\
    \nustar\ & 80802317002 & 2022-10-14T02:12:57 & 49624 \\
    \nustar\ & 80802317004 & 2022-10-26T08:35:11 & 46266\\
    \hline
\tabularnewline
\end{tabular}
\end{table}
%%%%%%%%%%%%%%%% Extended Data Table 1: datasets %%%%%%%%%%%%%%%%%%%%%%%%

%%%%%%%%%%%%%%%% Extended Data Table 2: glitch parameters %%%%%%%%%%%%%%%%%%%%%%%%
\begin{table}
\centering
    \caption{Best fit spin parameters between 2022 October 12 and November 06. Timing information for the glitches at times $t_{g1}$ and $t_{g2}$, \revi{inter-glitch and post-glitch frequency derivative, and frequency shift between two glitches are also included}.}
    \label{tab:timing_solution}
    \begin{tabular}{ll}
    \hline
    \hline
    Parameter & Value \\
    \hline
    Epoch (MJD) & 59865.6782217\\
    $\nu$ (Hz) & $0.3075277(6)$\\
    $\dot{\nu}$ (Hz s$^{-1}$) & $-1.7(6)\times10^{-11}$\\
    $t_{g1}$ (MJD) & $59866.63_{-0.02}^{+0.01}$\\
    $\Delta\nu_1$ (Hz) & $3.0(3) \times10^{-5}$\\
    $\Delta\dot{\nu}_1$ (Hz s$^{-1}$) & $-1.5(3)\times10^{-9}$\\
    \revi{$\dot{\nu}_1^*$ (Hz s$^{-1}$)} & \revi{$-1.5(3)\times10^{-9}$}\\
    $t_{g2}$ (MJD) & $59866.99(2)$\\
    $\Delta\nu_2$ (Hz) & $1.9(3) \times10^{-5}$\\
    $\Delta\dot{\nu}_2$ (Hz s$^{-1}$) & $1.5(3)\times10^{-9}$\\
    \revi{$\dot{\nu}_2^\dagger$ (Hz s$^{-1}$)} & \revi{$-4.9(1)\times10^{-12}$}\\
    $\Delta\nu_{12}^\ddagger$ (Hz) & $3.9(4)\times10^{-5}$\\
    $\chi^2/$dof & 83.2/64\\
    \hline
    \multicolumn{2}{l}{$^*$ inter-glitch frequency derivative}\\
    \multicolumn{2}{l}{$^{\dagger}$ post-glitch frequency derivative}\\
    \multicolumn{2}{l}{$^{\ddagger}$ frequency shift between two glitches}
\tabularnewline
\end{tabular}
\end{table}
%\mgbc{How about we include $\Delta\nu_{12}$, the frequency shift between the glitches?}
%%%%%%%%%%%%%%%% Extended Data Table 2: glitch parameters  %%%%%%%%%%%%%%%%%%%%%%%%

%%%%%%%%%%%%%%%% Extended Data Table 3: spectral parameters  %%%%%%%%%%%%%%%%%%%%%%%%
\begin{landscape}
\begin{table}
\centering
    \caption{Information of observations at each epoch around the glitches}
    \footnotesize
    \label{tab:observation_infomation}
    \begin{tabular}{c|ccccc}
    \hline
    \hline
    Epoch & Start/End time (\nicer\ MET) & \multicolumn{2}{c}{Exposure (s)} & \multicolumn{2}{c}{Count Rate (counts s$^{-1}$)}\\
    & & \nicer & \nustar & \nicer & \nustar  \\
    & & & FPMA/FPMB & & FPMA/FPMB\\
    \hline
    Quiescent & 214350080/214355957 & 1349 & 18580/18460 & $(4.9\pm{0.2})\times10^{-1}$ &$(2.1\pm{0.1})\times10^{-2}$/$(2.6\pm{0.1})\times10^{-2}$  \\
    A & 277061890/277225150 & 2034 & 2129/2111 & $1.27\pm{0.03}$ & (1.58$\pm{0.09})\times10^{-1}$/(1.47$\pm{0.09})\times10^{-1}$  \\  
    B & 277225150/277232029 & -- & 3166/3137 & -- & (2.02$\pm{0.08})\times10^{-1}$/(1.79$\pm{0.07})\times10^{-1}$  \\ 
    C & 277232029/277240909 & 1455 & 67/67 & 13.5$\pm{0.1}$ & 4.8$\pm{0.3}$/4.2$\pm{0.3}$  \\
    D & 277240909/277256736 & 377 & 490/485 & 3.7$\pm{0.1}$ & (2.8$\pm{0.2})\times10^{-1}$/(3.4$\pm{0.3})\times10^{-1}$  \\
    E & 277256736/277274900 & 1593 & 2101/2080 & 2.33$\pm{0.05}$ & (2.15$\pm{0.01})\times10^{-1}$/(2.36$\pm{0.01})\times10^{-1}$  \\
    F & 278238215/278288601 & 6592 & 46270/45850 & (7.89$\pm{0.01})\times10^{-1}$ & (9.39$\pm{0.02})\times10^{-2}$/(9.42$\pm{0.01})\times10^{-2}$  \\
    Burst & 277225150/277256736 & 94 & 824/823 & (4.90$\pm{0.02})\times10^{2}$ & (3.96$\pm{0.02})\times10$/(3.61$\pm{0.02})\times10$  \\
    \hline
\end{tabular}
\end{table}

\begin{table}
\centering
    \caption{Best-fit spectral parameters of the persistent X-ray emission}
    \footnotesize
    \label{tab:best_fit_spectral_parameter}
    \begin{tabular}{c|cccccc}
    \hline
    \hline
    Epoch & Absorbed/Unabsorbed flux $^{\rm{a}}$ & $N_{\rm{H}}$ & $kT$ & $R$ @6.6~kpc & $\Gamma$ & spectral model$^c$ \\
    & $10^{-11}$\,erg~s$^{-1}$~cm$^{-2}$ &  $10^{22}~\rm{cm^{-2}}$ & keV & km & & \\
    \hline
    \multicolumn{7}{c}{3C50 Background model for NICER data} \\
    \hline
    Quiescent & $0.49\pm{0.05}$ /$0.52\pm{0.05}$ & $2.1^{+0.5}_{-0.4}$ & $0.42^{+0.04}_{-0.03}$ & $1.8^{+0.5}_{-0.3}$ & $1.2\pm{0.1}$ & BB+PL \\
    A & $3.1^{+0.2}_{-0.3}$/$3.3\pm{0.3}$ & $2.6\pm{0.2}$ & $0.51\pm{0.03}$ & $2.1\pm{0.2}$ & $1.1\pm{0.1}$ & BB+PL \\
    B & $3.3^{+0.3}_{-0.4}$/$4.1\pm{0.3}$ & 2.4 (fixed) & 0.55 (fixed) & $1.9\pm{0.2}$ & 1.2$\pm{0.1}$ & BB+PL \\
    C & $93^{+4}_{-5}$/$95\pm{4}$ & $3.13^{+0.06}_{-0.07}$ & -- & -- & $1.22^{+0.03}_{-0.02}$ & PL \\
    D & $5.4^{+0.9}_{-1.1}$/$5.8^{+0.9}_{-0.6}$ & $2.4^{+0.3}_{-0.4}$ & $0.68\pm{+0.05}$ & $1.8^{+0.3}_{-0.2}$ & $1.6^{+0.2}_{-0.3}$ & BB+PL\\
    E & $3.1\pm{0.2}$/$3.3\pm{0.2}$ & $2.5^{+0.2}_{-0.1}$ & $0.61\pm{0.05}$ & $1.3^{+0.3}_{-0.2}$ & 1.8$\pm{0.1}$ & BB+PL \\
    F & $1.56^{+0.03}_{-0.05}$/$1.67\pm{0.03}$ & $3.4\pm{0.2}$ & $0.34\pm{0.02}$ & $4.4^{+0.3}_{-0.2}$ & $1.53\pm{0.2}$ & BB+PL\\
    Burst & $488\pm{2}$ / $525\pm{2}$ & $4.2^{+0.2}_{-0.1}$ & $0.17\pm{0.1}$ & $3.2^{+0.3}_{-0.2}\times10^2$ & $1.40\pm{0.02}^{\rm{b}}$ & BB+CPL \\
    \hline
    \multicolumn{7}{c}{SCORPEON Background model for NICER data} \\
    \hline
%    Quiescent & $0.49\pm{0.05}$ /$0.52\pm{0.05}$ & $2.1^{+0.5}_{-0.4}$ & $0.42^{+0.04}_{-0.03}$ & $1.8^{+0.5}_{-0.3}$ & $1.2\pm{0.1}$ & BB+PL \\
    A & $4.0^{+0.03}_{-0.03}$ / $4.1^\pm{0.3}$ & $1.47^{+0.06}_{-0.10}$ & $0.62^{+0.03}_{-0.02}$ & $1.48\pm{0.12}$ & $1.20^{+0.09}_{-0.10}$ & BB+PL \\
    B & $3.3^{+0.3}_{-0.4}$ / $4.1\pm{0.3}$ & 2.4 (fixed) & 0.55 (fixed) & $1.9\pm{0.2}$ & 1.2$\pm{0.1}$ & BB+PL \\
    C & $75^{+5}_{-5}$ / $77\pm{4}$ & $2.21^{+0.05}_{-0.06}$ & -- & -- & $1.33\pm{0.03}$ & PL \\
    D & $4.8^{+0.4}_{-0.5}$ / $5.3\pm{0.4}$ & $2.31^{+0.20}_{-0.19}$ & $0.54^{0.07}_{-0.08}$ & $2.2^{+1.0}_{-0.5}$  & $1.91\pm0.14$ & BB+PL\\
    E & $3.9\pm{0.2}$ / $4.2\pm{0.2}$ & $1.92^{+0.12}_{-0.11}$ & $0.56\pm{0.04}$ & $1.77^{+0.33}_{-0.21}$ & 1.65$\pm{0.08}$ & BB+PL \\
%    F & $1.56^{+0.03}_{-0.05}$ / $1.67\pm{0.03}$ & $3.4\pm{0.2}$ & $0.34\pm{0.02}$ & $4.4^{+0.3}_{-0.2}$ & $1.53\pm{0.2}$ & BB+PL\\
    \multicolumn{7}{l}{$^{\rm{a}}$ Fluxes are calculated in the energy range of 2--70 keV except for absorbed flux of epoch B and burst (10--70 keV).}\\
    \multicolumn{7}{l}{$^{\rm{b}}$ The photon index of the power-law with an  exponential rolloff at 37 keV.}\\
    \multicolumn{7}{l}{$^{\rm{c}}$ Corresponding \texttt{Xspec} models are, BB: \texttt{bbodyrad}, PL: \texttt{pegpwrlw}, and CPL: \texttt{cutoffpl}}\\
    %\multicolumn{7}{l}{*The denoted errors are statistical 1$\sigma$.}
    \end{tabular}
\end{table}
\end{landscape}
%%%%%%%%%%%%%%% Extended Data Figure 1: TOA Analysis %%%%%%%%%%%%%%%%%%%%%%%%

%%%%%%%%%%%%%%% Extended Data Figure 3: Corner Plot %%%%%%%%%%%%%%%%%%%%%%%%
%%%%%%%%%%%%%%% Extended Data Figure 3: Corner Plot %%%%%%%%%%%%%%%%%%%%%%%%

%%%%%%%%%%%%%%% Extended Data Figure 2: TOA Analysis reconfirm %%%%%%%%%%%%%%%%%%%%%%%%
%%%%%%%%%%%%%%% Extended Data Figure 2: TOA Analysis reconfirm%%%%%%%%%%%%%%%%%%%%%%%%

%%%%%%%%%%%%%%% Extended Data Figure 4: Energy Resolved Pulse Profile %%%%%%%%%%%%%%%%%%%%%%%%
%%%%%%%%%%%%%%% Extended Data Figure 4: Energy Resolved Pulse Profile %%%%%%%%%%%%%%%%%%%%%%%%

%%%%%%%%%%%%%%% Extended Data Figure 5: Time-Resolved Pulse Profile %%%%%%%%%%%%%%%%%%%%%%%%
%
%%%%%%%%%%%%%%% Extended Data Figure 5: Time-Resolved Pulse Profile %%%%%%%%%%%%%%%%%%%%%%%%

%%%%%%%%%%%%%%% Extended Data Figure 6: Persistent Spectral figures %%%%%%%%%%%%%%%%%%%%%%%%
%\begin{figure}
%\includegraphics[width=1.0\textwidth]{figure/spectral_fitting_results_persistent_20230331.pdf}
%\caption{Spectral fitting results of the persistent X-ray emission at each epoch, extracted from \nicer~(black), \nustar~FPMA(red), and \nustar~FPMB (green). Epoch labels are shown in %each panel. Left and right panels show the spectra before and after the FRB, respectively. At each epoch, raw count spectra (top panels) and fitting residuals (bottom panels) are %shown. Spectra are fitted by a blackbody plus power-law model except for the spectra of epoch C, at which a power-law model is used.\label{fig:spectral_fit_results_persistent}}
%\end{figure}
%
%%%%%%%%%%%%%%% Extended Data Figure 6: Persistent Spectral figures %%%%%%%%%%%%%%%%%%%%%%%%

%%%%%%%%%%%%%%% Extended Data Figure 7: Burst spectral figures %%%%%%%%%%%%%%%%%%%%%%%%
%
%%%%%%%%%%%%%%% Extended Data Figure 7: Burst spectral figures %%%%%%%%%%%%%%%%%%%%%%%%

\clearpage

%\begin{appendices}

%\section{Section title of first appendix}\label{secA1}

%An appendix contains supplementary information that is not an essential part of the text itself but which may be helpful in providing a more comprehensive understanding of the research problem or it is information that is too cumbersome to be included in the body of the paper.




%\end{appendices}

\clearpage
%\bibliography{reference_sgr1935}% common bib file


\end{document}